%% file: paper.tex
\documentclass[manuscript]{acmart}
\usepackage{geometry}
\usepackage{booktabs}
\usepackage{rotating}
\usepackage{pdflscape}
\usepackage[most]{tcolorbox}
\definecolor{medits}{RGB}{0, 0, 0}
\definecolor{edits}{RGB}{0,0,0}
\usepackage{soul}
\usepackage{array}

\AtBeginDocument{%
  }

\setcopyright{acmlicensed}
\copyrightyear{2024}
\acmYear{2024}
\acmDOI{XXXXXXX.XXXXXXX}

\acmConference[CSCW '25]{The 28th ACM SIGCHI Conference on Computer-Supported Cooperative Work \& Social Computing}{Oct -- Nov,
  2025}{Location Pending}
\acmISBN{978-1-4503-XXXX-X/18/06}

\begin{document}

\title{Building Proactive and Instant-Reactive Safety Designs
  to Address Harassment in Social Virtual Reality}


\author{Zhehui Liao}
\affiliation{%
  \institution{University of Washington}
  \city{Seattle}
  \country{US}
}

\author{Hanwen Zhao}
\affiliation{%
  \institution{University of Washington}
  \city{Seattle}
  \country{US}
}

\author{Ayush Kulkarni}
\affiliation{%
  \institution{University of Washington}
  \city{Seattle}
  \country{US}
}

\author{Shaan Singh Chattrath}
\affiliation{%
  \institution{University of Washington}
  \city{Seattle}
  \country{US}
}

\author{Amy X. Zhang}
\affiliation{%
  \institution{University of Washington}
  \city{Seattle}
  \country{US}
}

\renewcommand{\shortauthors}{}

\begin{abstract}
 \input{0_abstract}
\end{abstract}

\maketitle
\input{1_intro}

\input{2_related}

\input{3_formative}

\input{4_system}

\input{5_evaluation}

\input{6_results}

\input{7_discussion}

\input{8_conclusion}

\bibliographystyle{ACM-Reference-Format}
\bibliography{8_reference}
\input{9_appendix}

\end{document}

%% file: 0_abstract.tex
Social Virtual Reality (VR) games offer immersive socialization experiences but pose significant challenges of harassment. Common solutions, such as reporting and moderation, address harassment after it happens but fail to prevent or stop harassment in the moment.
In this study, we explore and design proactive and instant-reactive safety designs to mitigate harassment in social VR. Proactive designs prevent harassment from occurring, while instant-reactive designs minimize harm during incidents. We explore three directions for design: user-initiated personal bubbles, clarifying social norms, and encouraging bystander intervention. Through an iterative process, we first conducted a formative interview study to determine design goals for making these features effective, fit user needs, and robust to manipulation. We then implemented Puffer, an integrated safety system that includes a suite of proactive and instant-reactive features, as a social VR prototype. 
From an evaluation using simulated scenarios with participants, we find evidence that Puffer can help protect players during emergencies, foster prosocial norms, and create more positive social interactions. 
We conclude by discussing how system safety features can be designed to complement existing proactive and instant-reactive strategies, particularly for people with marginalized identities.


%% file: 1_intro.tex
\section{Introduction}

\emph{\textbf{Content Warning:} This paper includes discussions about topics such as sexual harassment and identity-based harassment.}

In recent years, virtual reality (VR) technology~\cite{web} and socializing in virtual worlds via VR headsets~\cite{Maloney} have become more widespread.
Researchers find benefits to the immersion of social VR, including 
facilitating long-distance relationships, older adults' communication, and virtual intimacy \cite{Zamanifard, Baker, aggeler2022, Freeman2021}. 
However, the immersive nature of VR also presents unique challenges for player safety. Multiple reported harassment cases highlight the severity and negative impacts of harassment in VR. For instance, shortly after the release of Horizon Worlds, Meta's social VR platform, beta testers and researchers experienced sexual harassment from strangers, including misogynistic comments, groping, and rape, indicating the inadequacy of current safety protocols \cite{sparks, basu, soon}. Indeed, researchers found harassment to be a prevalent and significant problem, with 42\% of social VR players witnessing it at least once \cite{Shriram}. Even worse, harassment in social VR, similar to other online platforms, has disparate impacts on \textcolor{edits}{players who are marginalized in social VR gaming due to identity-based harassment, including} women, LGBTQ+ populations, and ethnic minority players~\cite{Freeman2022, womenVR}. 


Many defenses against harassment in online social spaces focus on \textit{reactive} approaches, which respond to harassment after it happens to reduce long-term repercussions. For instance, researchers have worked on improving the effectiveness of moderation \cite{filterbuddy, Jhaver2023PersonalizingCM, Jhaver2019HumanMachineCF} or the reporting experience \cite{Abhinaya, bystander3}. 
However, reactive approaches alone are insufficient for the context of social VR. First, evidence of harassment is harder to collect in social VR, where historical records of interactions are sparse, placing a greater burden on victims to make their case~\cite{Abhinaya}. Second, the current lack of moderators and the lack of trust between moderators and players pose challenges to effectively moderating social VR spaces \cite{vrmod}. Finally, the immersive nature of VR makes harassment experiences
more realistic and traumatizing compared to other online platforms \cite{Freeman2022}. Even if reactive strategies such as reporting could mitigate long-term consequences, the harm experienced could be hard to remediate.

In this work, we explore the design space of \textit{proactive} and \textit{instant-reactive} safety features for social VR.
Similar to strategies to combat sexual harassment in the workplace~\cite{McDonald2015}, we distinguish ``proactive strategies'' that aim to \textit{prevent future harm} and ``instant-reactive'' strategies that aim to \textit{stop further harm at the moment} from ``delayed-reactive strategies'' that mitigate harm some time after it has occurred.
\textcolor{edits}
{Prior researchers found that proactive strategies, despite their potential to prevent harm, are less common in intervention systems for online games \cite{tametoxicplayer}.}
Given the unique affordances of VR, we focus on features that protect against \textit{embodied harassment}, or the violation of other players' avatars and personal space, including some types of sexual harassment, such as groping and rape, violent behaviors, such as throwing objects at a player, and even unintentional actions, such as standing too close to a player.

\begin{figure}
  \centering
  \includegraphics[width=1.0\textwidth]{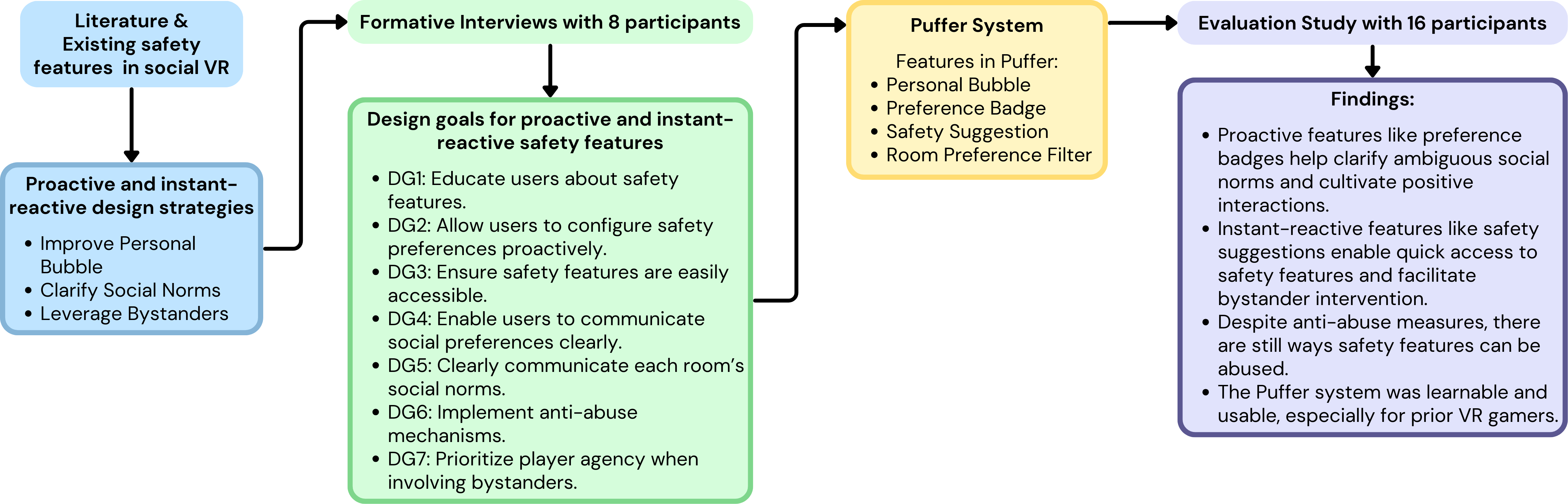}
  \caption{\textcolor{edits}{Overview of the study process and findings. We took an iterative approach to design proactive and instant-reactive safety features in Puffer and incorporated participant feedback at different stages of the study.}} 
  \label{fig:method}
\end{figure}

\textcolor{edits}{As shown in Figure~\ref{fig:method}}, we started from design and literature exploration and from there chose to investigate three kinds of proactive and instant-reactive strategies that we considered most promising: \textcolor{medits}{improving ``personal bubble'' designs, clarifying social norms, and leveraging bystanders.}
From a formative interview with 8 gamers to explore their reactions to these strategies, 
we found that \textcolor{medits}{participants prioritized preventing or stopping harassment but were unaware of or unable to access existing safety features. Participants also needed clear ways to share their preferences and better understand social norms in game rooms. Additionally, some women and queer players were concerned about the misuse of safety features by malicious gamers.} \textcolor{edits}{We distill these findings into 7 design goals for proactive and instant-reactive safety features in VR (Figure~\ref{fig:method})}.

We then built Puffer, a social VR safety system that implements proactive and instant-reactive designs against harassment into a holistic, interconnected system, \textcolor{edits}{considering our design goals from the formative study}. \textit{Proactive features} include (1) pre-configuring a personal bubble along several dimensions and (2) setting norm expectations using both personal badges
and filterable room characteristics.
\textit{Instant-reactive features} include (1) safety suggestions that a bystander can trigger and \textcolor{medits}{(2) a shortcut to toggle a personal bubble; promoting players to help each other and other quick access to safety features.} Puffer is also built with mitigations for potential abuse in response to participants' concerns, such as cool-down periods and ways for players to opt-out, both intended to avoid conflicts with players' existing coping strategies. 

We invited 16 participants to paired in-person evaluation sessions where they tested Puffer \textcolor{medits}{through tasks and simulated scenarios.} 
Overall, we found that Puffer helped protect user safety both proactively and instant-reactively. Proactive features 
clarify the ambiguous social context in the roleplay scenarios, helping participants avoid unintentional conflicts. Instant-reactive features provided immediate support that participants quickly accessed to protect themselves or others. 
\textcolor{medits}{Beyond combating} embodied harassment, participants envisioned these features collectively creating opportunities for more positive gaming experiences. For example, participants suggested these features could foster mutual support by encouraging players to look out for one another and improve socialization by helping them find like-minded players and suitable game worlds. They also shared concerns about bad actors misusing the system, such as spamming or intentionally violating others' preferences.
We conclude with a discussion of how proactive and instant-reactive designs can complement marginalized players' existing strategies and be customizable and voluntary given diverse safety needs.

%% file: 2_related.tex
\section{Related Work}
\subsection {Online Harassment and Strategies for Mitigating its Impact}

Online harassment encompasses a wide range of abusive behaviors, including public shaming, doxing, and threats. 
Extensive HCI research examines online harassment and solutions to address it \cite{vogels2021state, Vitak}, identifying problems such as inconsistent and ambiguous platform policies \cite{Pater} and insufficiency of current technology to label and mitigate it \cite{blackwell2017}. 
In the realm of online gaming, infamous for harassment, researchers identified the lack of social support and safety tools as well as the normalization of toxicity in games \cite{Frommel}. Despite the increasing diversity in gamer identity, women and LGBTQ+ gamers remain significantly exposed to identity-based harassment \cite{adl2022}. Facing an unfriendly gaming environment and lack of support from platforms, women gamers have developed unique strategies to avoid or protect themselves from harassment, but they still experience more emotional distress and higher risks than men \cite{coteWomenGamer, Fox}.

Regarding workplace sexual harassment, researchers in violence prevention developed frameworks that categorize organizational strategies to combat sexual harassment into primary prevention, secondary intervention, and tertiary intervention \cite{McDonald2015}. Primary prevention strategies are \textbf{proactive}, bringing changes to the environment and creating systematic and clear norms and measures to prevent harassment. Secondary and tertiary interventions are both \textbf{reactive}, mitigating the harm, negotiating its impacts, \textcolor{edits}{and bringing a sense of fairness to victims of harassment.} We characterize secondary interventions as ``\textbf{instant-reactive}'' since they prioritize immediately stopping the ongoing harassment to minimize its impact while tertiary interventions can be considered ``\textbf{delayed-reactive}'' since they focus on dealing with long-term consequences. 

Strategies and systems developed by HCI researchers to combat online harassment can be divided into proactive and reactive, as well. 
When designing proactive features, 
researchers have re-envisioned social platforms through the feminist lens of affirmative consent, prioritizing user agency and control in preventing harassment \cite{Im}. Researchers also found that modeling and clarifying social norms help prevent harassment, protect users' privacy and safety, and encourage prosocial behavior in online communities \cite{norm_vulnerability, Matias2019PreventingHA,10.1145/3686929}. 


In terms of instant-reactive strategies, 
HCI researchers have reaffirmed the benefits, \textcolor{edits}{identified in school and workplace analyses \cite{McMahon2012WhenCI, Mainwaring2022ASR},} of bystander intervention in reducing perceptions of justified harassment \cite{Blackwell2018WhenOH} and have designed ways to integrate it into online platforms to combat cyberbullying \cite{bystander1, bystander2, bystander3}. 
Finally, significant HCI work has focused on delayed-reactive strategies, notably tools for moderating, documenting, and reporting the harassment or harasser, giving the targeted person support or amends or limiting their exposure to harm, and community-wide governance mechanisms for justice and mediation~\cite{amy, Sultana2021UnmochonAT,restorative,respect,Seering2019ModeratorEA,wang2023reporting,filterbuddy}.


\subsection{Challenges of Harassment in Social VR}


Although the immersive and realistic nature of VR can produce many benefits \cite{Baker, Zamanifard, aggeler2022, Freeman2021, avatar1, avatar2}, it also introduces unique challenges that make harassment more traumatizing. Indeed, researchers identified multiple 

emerging safety risks unique to social VR due to the complex social interaction dynamics and multi-modal virtual simulations \cite{Zheng2023}. Harassment is also found to be a major problem, with 42\% of social VR players witnessing it at least once \cite{Shriram}. 
This problem received significant attention after players like Jordan Belamire publicly shared their experiences of being groped \cite{belamire}; researchers studying the Reddit discussions about this incident found that it exposed the lack of consensus on what constitutes harassment among players \cite{Sparrow} and identified challenges for VR platform governance \cite{Lindsay}. 

In fact, the ambiguous social norms of what constitutes harassment versus the ``fun/play'' behaviors in social VR exacerbate embodied harassment and pose challenges to regulation and building safer environments \cite{Lindsay, Freeman2022, Zheng2023}. For example, Zheng's team found attackers often viewed harassment as ``just a game,'' where offline norms should not apply, and the co-existence of minors and adults creates tensions that could escalate to harassment due to the different perceptions and behaviors of these two demographics \cite{Zheng2023}. Facing these challenges, Blackwell et al.~proposed a bottom-up approach, which involves fostering positive social norms within VR communities to regulate user behavior, as opposed to traditional top-down moderation \cite{Lindsay}. Additional research emphasized the importance of fostering transparent social norms in VR environments since they provide vital cues that help players understand appropriate ways to interact with other players and employ safety features \cite{Freeman2022}. Researchers also conducted studies with industry experts and VR developers to suggest design considerations for pro-social interactions and identify difficulties from the implementation and development side \cite{McVeigh, Abhinaya}. 

Similar to other online social platforms, players with marginalized identities, such as women, LGBTQ+ populations, and ethnic minorities, face higher risks of experiencing harassment targeting their identities in social VR \cite{Freeman2022}. To protect themselves, women players adopt both system safety tools and develop protective  strategies.
This research not only calls for more safety designs to address such harassment but reminds us to consider marginalized players' strategies and mindsets when creating new safety designs.

\subsection{Proactive and Reactive Safety Strategies in Social VR}

Due to the prevalence and severity of harassment in social VR, game platforms and researchers have implemented and studied various proactive and reactive strategies.
Existing safety features in social VR games predominantly embrace delayed-reactive approaches \cite{Freeman2022, Zheng2023, birdcage}, such as user-initiated blocking, muting, reporting, and moderator-initiated sanctions.
However, delayed-reactive strategies, such as moderation and reporting, have significant drawbacks in social VR.
While moderation is a major platform-sanctioned strategy, 
researchers found the lack of moderators and moderators not knowing the necessary context \cite{vrmod} render this approach non-optimal for harassment. 
They also identified concerns about players' lack of trust in moderators and moderation biases \cite{Lindsay, Freeman2022}; as a result, researchers have considered adopting AI for social VR moderation \cite{AIMod}. Further, prior work on reporting identified difficulties in collecting evidence because of the ephemeral nature of incidents and the cumbersome reporting process \cite{Freeman2022, Abhinaya}. 
Finally, due to the immersive experience of social VR,
players find embodied harassment to be more traumatizing and harder to remedy via delayed reactive strategies \cite{Freeman2022}.

Due to challenges of applying delayed reactive strategies and the harmful nature of embodied harassment, many platforms adopt proactive strategies. 
One example is the personal bubble, a safety tool designed to combat embodied harassment that is implemented in most major social VR games. Bubbles guard players' personal space, and, when activated, prevent others from entering and touching player avatars. They can also be used as an instant-reactive tool for players to immediately engage when harassed. Despite being praised by many players, bubbles have been criticized for blocking players from having a complete experience since virtual ``touch'' is considered an essential part of social VR \cite{Freeman2022}. 
In addition, the sheer range of implementations and differences in accessing the feature may confuse users further (see Appendix~\ref{apx:bubble} for a landscape analysis). Another proactive strategy from VRChat is the Trust System, used to measure the trustworthiness of each player. However, researchers found that the system may instead exacerbate discrimination and stereotypes towards low-ranked players, presenting challenges to new players \cite{vrchatTrust}. 

In addition to platform tools, women social VR gamers have developed their own proactive strategies to protect themselves. For instance, women players disguise their gender identity by avoiding feminine-presenting avatars or editing their voices to avoid unwanted attention. However, these approaches could contribute to stereotypes regarding the lack of women gamers \cite{womenVR}. Finally, recent research on proactive safety designs in social VR has focused on the lens of consent \cite{consensual, birdcage, dating}. Researchers re-envisioned safety designs to practice consent by setting boundaries in social VR, balancing protection and interaction.

%% file: 3_formative.tex
\section{Formative Study}

Based on the literature review and review of existing safety features in industry social VR products, we chose to investigate three proactive and instant-reactive strategies: user-initiated personal bubbles, clarifying social norms, and encouraging bystander intervention. \textcolor{edits}{We explored the design space of these three strategies through a formative interview study to develop design goals for implementation. Participants experienced variations of each strategy via playtesting in VR or reviewing design sketches while speaking aloud their reactions. We adopted this approach to leverage and build upon prior research findings, with the goal of the formative study to rapidly iterate and refine our strategies based on participants’ feedback.}

\subsection{Proactive and Instant-Reactive Safety Strategies}

\textbf{Personal Bubble}.
In a prior study, one-third of participants praised the personal bubble feature in RecRoom and VRChat as a strategy that best prevents harassment relative to strategies such as blocking \cite{Freeman2022}. However, we observed that personal bubbles also present some usability problems. 
For instance, when a personal bubble was turned on by default, new players thought their game had glitched \cite{vrchatPB, vrchatPB2}.
Depending on the platform, players are not informed of the effect of the personal bubble or must go through multiple steps to activate it (see Appendix \ref{apx:bubble} analysis). 
Therefore, we \textcolor{edits}{decided to investigate} potential ways to improve the usability of personal bubbles, including  (1) creating easier and quicker ways to activate and customize one's personal bubble, (2) designing bubbles to be more aware of the gaming context, such as when a player might want to use a bubble or not, and (3) communicating the effects of a personal bubble to players to avoid confusion. During the session, participants tried locating and using the personal bubble feature in RecRoom. Then, they reflected on that experience and envisioned ways to improve personal bubbles.

\textbf{Social Norms}.
Researchers have found that the lack of social norms in VR, such as expectations for how to interact with one another, is a major reason for perceived harassment  \cite{Lindsay}. Social VR players shared that they were often unsure about how to approach another stranger \cite{Freeman2022}. Ambiguous social norms blur the boundary between harassment and inappropriate ``fun/play'' behaviors, leading to unintentional violations of others' autonomy \cite{Lindsay, Freeman2022, Zheng2023}.
Thus, we explored how to foster a norm of communicating and respecting other players' preferences in social VR through a self-selected preference badge system.
Our designs for the formative study included two categories of customizable preference badges: (1) a ``behavior'' badge that defines a player's comfort level in their interaction distance from other players, with options of Close, Medium, and Distanced, and (2) a ``sound'' badge defines a player's comfort level with the volume of their environment or interaction with others, with options of Quiet, Medium, and Loud.
Badges include an icon and label and are shown above a player avatar's head. Players can also add comments to their badge's description, which other players can view by clicking on the badge. We envisioned that these preferences could also help players find game rooms with norms suited to them. For instance, a player with a ``Quiet'' preference could filter out louder worlds. In current social VR games, room labels exist to create filters for players browsing all available rooms, such as ``Hot,'' ``New,'' and ``\#horror'' in RecRoom; however, these labels mostly address player interests, room themes, and room popularity but fail to share social norms in the room.

\textbf{Bystander Intervention.}
While moderators play a crucial role in maintaining safety in social VR~\cite{vrmod, Lindsay}, bystanders are more readily available for faster intervention, \textcolor{edits}{and intervention systems enacted by players were identified as a promising way to create positive, empowering effects on player behavior \cite{tametoxicplayer}. }
Teaching players to be supportive bystanders also enlists them as contributors to a healthy environment and communicates social norms in game\cite{bystander2}.

Though HCI researchers have studied encouraging bystander intervention on social media platforms~\cite{bystander1, bystander2, bystander3}, this strategy has not been studied in the multiplayer gaming and social VR context.
Thus, we solicited feedback on three designs for bystander intervention in social VR; all share the same scenario, where bystander B encounters player P, who is being visibly harassed by a group of players. Our designs include: (1) B sending a private message to P to invite them to a shared Safe Zone in which only B and P can interact with each other, but no other players can virtually touch or talk to them, (2) B sending a private message to P to remind them of safety features that could be helpful for P's situation, e.g., B sending a message to share the Block feature and if P accepts, the system will automatically activate the Block feature for P, and (3) B opening the list of players, and, if a moderator is in this room, sending a request for assistance. 

\begin{table}
\footnotesize 
\centering
\begin{tabular}{p{0.3cm} p{0.7cm} p{1cm} p{2.6cm} p{3.7cm} p{4.6cm}} 
\toprule
 & Age & Gender & Race & Sexuality & VR games played  \\ \midrule
P1 & 18-22 & Man & Asian & Heterosexual & Beat Saber\\
P2 & 18-22 & Man & Hispanic or Latino, White & Prefer Not to Disclose & Echo , Pavlov, VRChat, Poker Stars\\ 
P3 & 18-22 & Woman & Asian, White & Queer & Superhot\\
P4 & 18-22 & Woman & Asian & Heterosexual & Beat Saber, Rec Room, Apex Construct, Echo VR, Superhot, Gorilla Tag \\
P5 & 18-22 & Man & White & Heterosexual & Beat Saber, Rec Room, Gorilla Tag, Warplanes, Echo VR\\
P6 & 18-22 & Man & Asian & Heterosexual & Half life alyx, Superhot, Beat Saber\\
P7 & 28-32 & Woman & Asian, White & Bisexual, Asexual/aromantic/agender & N/A (First-player shooting gamer) \\
P8 & 18-22 & Woman & Asian & Heterosexual & N/A (MOBA gamer)\\
\bottomrule
\end{tabular}
\caption{Formative Study's Participant Demographics.}
\label{tab:formative}
\end{table}

\subsection{Participants and Methods}
To develop design goals for proactive and instant-reactive safety features in social VR, we first conducted an interview study with 8 social VR and general gamers (Table~\ref{tab:demographics}) to learn from their experiences and elicit feedback on our design ideas. Specifically, through posts and messages in college student groups and gaming communities and snowballing, we recruited 8 participants for a 60-minute semi-structured interview session, prioritizing women and LGBTQ+ participants as these demographics are more likely to experience harassment \cite{Freeman2022}. We also included participants with varied levels of social VR experiences. 


Interviews were conducted in person at a university campus during June 2023. We first asked about participants' previous experiences with safety features and harassment in multi-player games and social VR.
\textcolor{edits}{Next, we asked for feedback on each safety strategy described in Section 3.1. For the personal bubble strategy, participants tried the bubble when playing RecRoom with an Oculus Quest2 headset and discussed what they found helpful versus difficult. 
We chose RecRoom instead of other social VR games for bubble testing because it had two bubble implementations, ``personal space'' and ``ignore bubble,'' which helped us understand participants' preference for the bubble effect \cite{recroom_pb}. For the other two strategies, clarifying social norms and encouraging bystander intervention, we sought feedback by asking participants to evaluate and iterate on low-fidelity design sketches (see sketches in Appendix~\ref{apx:sketches}). We focused on identifying scenarios in which participants would and would not leverage these strategies and areas of improvement to ensure safety features met their needs.} 
Participants received a gift card of \$20 as compensation. 

\textcolor{edits}{We recorded and audio-transcribed the interview data; subsequently, the first author manually corrected the transcription. The first author then conducted an inductive thematic analysis to draw insights \cite{thematic_analysis, reflexive_thematic_analysis}, reading all interview data, marking related sections of data, and deriving initial codes. Then, she coded the related sections, discussed them with the other author, and adjusted the codes. Together, they generated themes by collating the codes and refining themes again through discussions. Appendix~\ref{formative_codebook}} presents the final codebook.

\subsection{Results}

We organized our findings into three central themes. 
These themes informed our seven design goals for improved safety features that enable proactively and instant-reactively responding to embodied harassment.

\subsubsection{Preventing harassment and stopping it instantly is the first priority, yet participants do not know about or cannot easily access safety features.}
Participants affirmed that they prioritized both proactive and instant-reactive measures relative to delayed-reactive ones. 
For example, P4 
shared that when gaming in social VR, she proactively mutes herself inside the game and opens an audio chat with friends via Discord so that strangers in the game do not hear her feminine-sounding voice.
However, her audio was on one occasion auto-turned on, and she was subsequently harassed and immediately quit the room. Recalling that situation, she noted that she would not want a moderator to mediate because waiting for one might prolong the harm. Quitting the room was her instant-reactive strategy for self-protection. 

Participants highlighted that safety features should be easy to access during an emergency to be effective.
Although multiple safety features exist in current social VR games (e.g., blocking, personal bubbles, safety zones, and reporting), many are difficult to quickly access when harassment occurs. In emergencies, players may be nervous, pressured, and error-prone, increasing the difficulty of navigating to the exact location of a safety feature. As P3 shared, ``\textit{...getting harassed...at that point, you want to be able to go to a safe space as soon as possible to change the settings. If it's not immediately brought up,  then people won't know where to go in order to do that.}'' 

Further, participants experienced difficulties locating and learning about existing safety features in the first place.
As of June 2023, social VR games such as VRChat and RecRoom 
have no introduction to their safety features or tutorials on how to use them.
P2, a former player of VRChat, shared that he never heard about personal bubbles until our research session. 
Some participants new to social VR also remarked on the lack of discoverability of the feature.
\textcolor{medits}{Even with access to feature information, participants, both new and experienced social VR players, struggled to understand the personal bubble's effect through text description alone. For example, they found the customizable bubble size differences unclear.}
These findings motivate our first set of design goals regarding social VR safety features.




\begin{center}
\begin{tcolorbox}[colframe=black, colback=white, boxrule=0.3mm, width=1\textwidth, arc=0mm, auto outer arc]

\textbf{DG1}: Users should be educated about the existence and impact of safety features, including when setting up preferences as a new user and in situations when they might be useful.

\textbf{DG2}: 
Users should be able to proactively configure safety preferences. Tools could then support users to more easily detect potentially unwanted interactions in order to avoid them. 

\textbf{DG3}: Safety features should be easily accessible when harassment happens to minimize harm. This includes providing shortcuts to safety features in an emergency.




\end{tcolorbox}
\end{center}

\subsubsection{Participants find value in making social norms more explicit and communicating their preferences to others, though they prefer methods with low overhead.}

Most participants did not communicate their preferences with others unless it was part of the game's goals, such as for team collaboration. 
The lack of communication was partly due to participants' lack of trust in others or their low expectations for others to respect their preferences.
These factors contribute to a broader lack of established norms in different game worlds and perceptions of a toxic environment, even as many players, including our participants, wish to interact in good faith and worry about being perceived as rude.
Part of the problem is the high overhead of communication and the possibility of misinterpretation.
Thus, participants liked the designs for preference badges regarding distance and sound.
Both P4 and P6 suggested an additional ``mood badge'' to indicate whether the user is in the mood to socialize since they occasionally logged on to sight-see or wait for friends and were not interested in engaging with new people. However, they worried that communicating their lack of interest would be misinterpreted as hostile, so their strategy was simply to ignore others. Instead, a badge could subtly communicate this preference without needing to engage in dialogue and without appearing rude by ignoring others. 

In addition, every participant was interested in the design idea involving matching their safety preferences to room social norms.
For instance, players who have a ``quiet'' preference could filter out rooms with ``loud'' labels, proactively preventing sound-based harm from occurring. 
P5, a player of RecRoom, shared that ``\textit{tagged servers…that's incredibly useful. I would like the game to have that sort of thing}.'' P5 also suggested that room labels should be voted on by players to be more accurate and a way for players to explicitly co-construct their desired social norms for the space.

\begin{center}
\begin{tcolorbox}[colframe=black, colback=white, boxrule=0.3mm, width=1\textwidth, arc=0mm, auto outer arc]

\textbf{DG4}: Users should be able to easily communicate their social preferences to nearby users. Such communication should be done in a way that is clear yet unobtrusive.

\textbf{DG5}: Rooms should more clearly communicate the social norms of the space so that users can proactively decide whether to enter.


\end{tcolorbox}
\end{center}
\subsubsection{Participants caution that safety features must be robust to misuse by malicious gamers.}

One major concern raised by participants was the misuse of safety features by malicious gamers to further harassment.
Participants of different genders showed distinct attitudes toward safety designs. 
P1, P2, and P6, who all identified as men, did not express a strong need for safety designs to protect themselves because they had become accustomed to the toxic environment. 
In comparison, women participants showed higher interest in adopting these safety designs but were concerned about potential abuse of them. 
P3, P4, and P7, in particular, had a zero-trust mindset for strangers in social VR and more generally in online multiplayer games, instead relying on themselves or their friends to manage potential risks and harassment.
As a result, they were concerned about the bystander intervention designs and questioned the intentions of bystanders.
For example, P4 shared an incident when strangers in Gorilla Tag touched her avatar without consent, and her first reaction was to quit the game. Had bystanders tried to help, P4 said she would ignore them unless they were friends, saying: ``\textit{I don't trust people who are bystanders to be good people. Like I could definitely see it...being abused}.'' 

Participants had the same concern about preference badges, worrying that players who shared their preferences would be targeted for harassment. For example, a player who wears a badge preferring quiet environments could be targeted by offenders shouting at them. As a woman and queer gamer, P3 shared that ``\textit{I do worry a little bit about putting the trust in the people to respect [the preference badge]},'' and suggested combining the preference badge with system controls. Suppose players indicate the preference to maintain a certain distance from other players, a personal bubble could be automatically enabled to enforce this distance. If players indicate preferences for a quiet environment, noise levels could be monitored and loud sounds filtered out automatically. In this way, their badges would both communicate and enforce their preference.



Among the three bystander intervention designs, intervening via chat received the most positive feedback
because the player maintains agency in deciding if they want to use the safety feature. 
In contrast, intervening physically requires a high-stakes ``physical'' interaction with a potentially untrusted bystander, and intervening via moderators involves moderators whom many participants complained about being too slow. Introducing a moderator also changes the issue from being victim-centered to offender-centered, where the moderator is considered the authority who metes out punishments, which could lead the player who experienced harassment to lose the power to make decisions for themselves. As P3 shared, ``\textit{Give [the players who experience harassment] the power to decide what they want to do and like, `Hey, remember, you have these options.' I think that's great.}''





\begin{center}
\begin{tcolorbox}[colframe=black, colback=white, boxrule=0.3mm, width=1\textwidth, arc=0mm, auto outer arc]

\textbf{DG6}: Safety features must consider and implement anti-abuse mechanisms to protect users who choose to turn on a feature from bad actors seeking to exploit it.

\textbf{DG7}: In determining how to involve third parties such as bystanders or moderators, safety features should center player agency and not make decisions on their behalf. 


\end{tcolorbox}
\end{center}


%% file: 4_system.tex
\section{Puffer: A Proactive and Instant-Reactive Social VR Safety System}

\begin{figure}
    \centering
    \includegraphics[width=.9\textwidth]{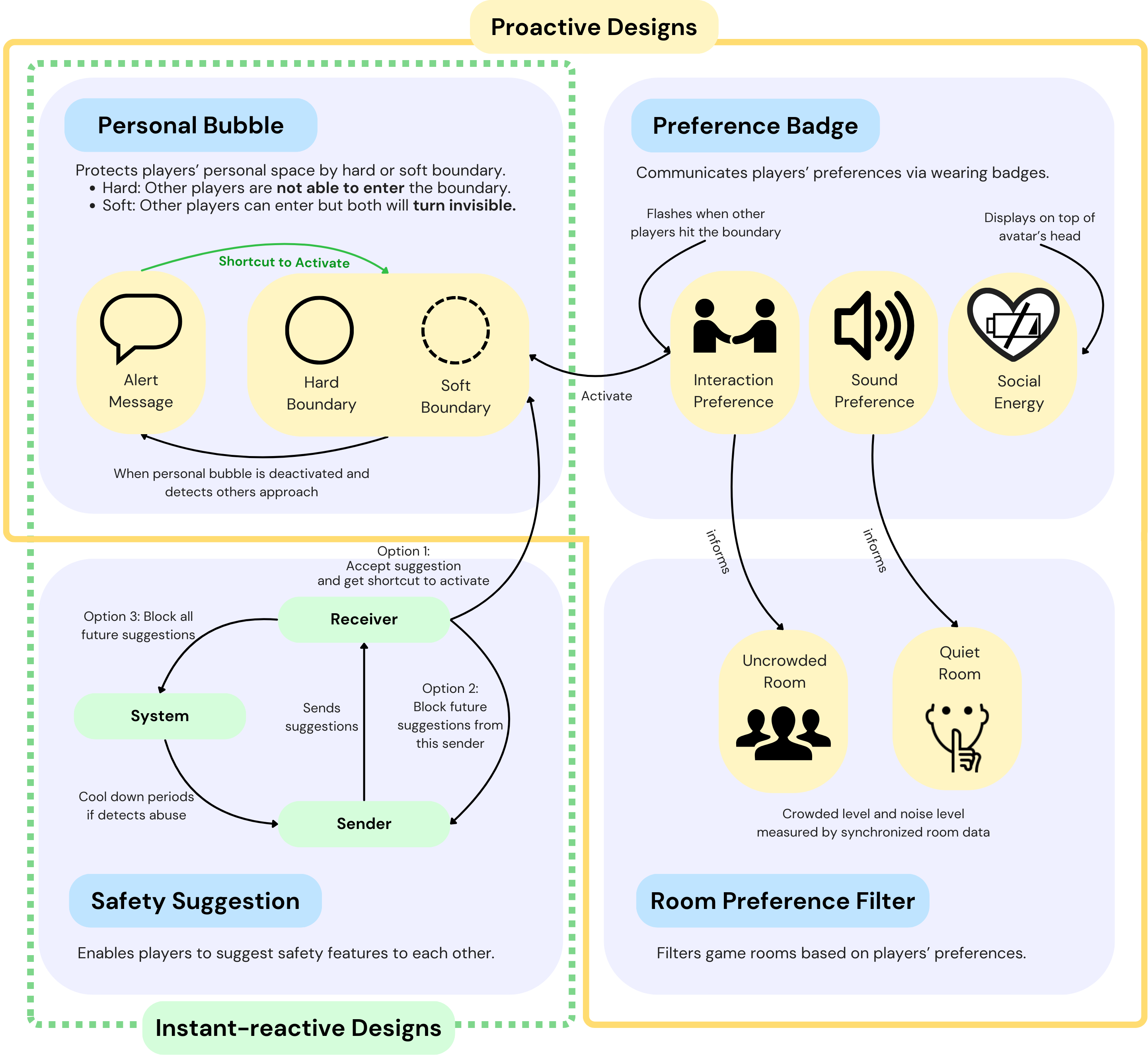}
    \caption{Puffer Overview. Proactive features, including the personal bubble, preference badges, and room preference filters, are grouped by yellow full lines; instant-reactive features, including the personal bubble and safety suggestions are grouped by green dotted lines.} 
    \label{fig:flowchart} 
\end{figure}

Building on the design goals from our formative study, we created Puffer, a social VR safety system with proactive and instant-reactive features to protect users \textcolor{edits}{from embodied harassment}.\footnote{We named our system Puffer after the pufferfish, animals that proactively protect themselves by puffing up with spikes, a clear indication of their preference, viz., `Please do not get close to me!''} Puffer offers four interconnected features: personal bubbles, preference badges, safety suggestions from bystanders, and room preference filters. 
\textcolor{edits}{Proactive features prevent harassment and clarify norms, while instant-reactive features provide prompt support initiated by players themselves or others. Personal bubbles improve upon existing social VR safety designs, and the other three do not exist in any social VR games (as of June 2023). Figure \ref{fig:flowchart} shows all Puffer features and how they interact.}


\begin{figure}
    \centering
    \includegraphics[width=1.0\textwidth]{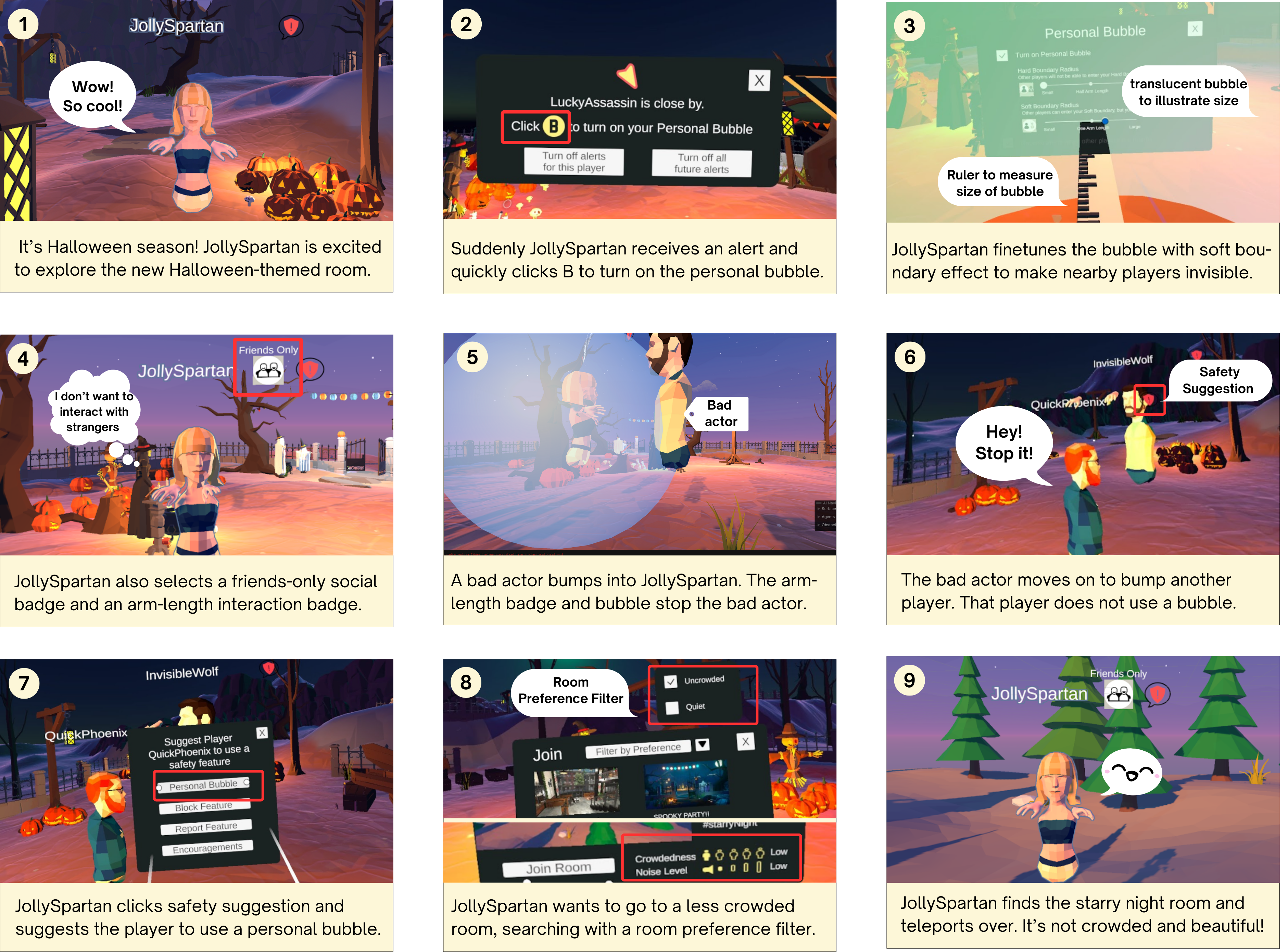}
    \caption{Storyboard: Player JollySpartan's journey with Puffer's safety features in social VR.} 
    \label{fig:sb}
\end{figure}



\subsection{User Scenario}
Figure \ref{fig:sb} \textcolor{medits}{presents a storyboard of a user scenario featuring JollySpartan, an experienced social VR player familiar with Puffer's safety features.}
In Figure~\ref{fig:sb}, 
JollySpartan, exploring a popular new game room (Panel 1), receives an alert about another player approaching from behind. Because JollySpartan has settings set to alert if a stranger gets too close, the system asks if they want to activate their personal bubble. By clicking on the controller's button B, JollySpartan successfully enables the bubble (Panel 2). 
JollySpartan then finetunes the \textbf{personal bubble} to use a soft boundary making nearby players invisible, as the crowded room makes it hard to see (Panel 3). Realizing they forgot to put on \textbf{preference badges}, JollySpartan adds the `friends-only' and `no physical interaction' badges to avoid socialization with strangers (Panel 4).

JollySpartan notices that another player, InvisibleWolf, is intentionally bumping into others (Panel 5). When InvisibleWolf tries to bump into JollySpartan, their personal bubble flashes to everyone to show their preference. Because of the soft boundary effect, InvisibleWolf and JollySpartan become invisible to each other when getting close, and InvisibleWolf thus cannot interact with JollySpartan's avatar (Panel 5). InvisibleWolf moves on to bump another player, who, lacking a personal bubble, seems annoyed (Panel 6). JollySpartan sends that player a \textbf{safety suggestion} that includes a shortcut to turn on the bubble (Panel 7). 

JollySpartan does not want to engage with InvisibleWolf and decides to find a less crowded room. On the room menu, the \textbf{room preference filter} is automatically set for uncrowded based on JollySpartan’s `no physical interaction' badge (Panel 8, top). Filtering to see all uncrowded rooms and finding one that matches their preference (Panel 8, bottom),  
JollySpartan enters the new room, which offers a more suitable vibe to enjoy (Panel 9).



\subsection{System Features}

\begin{figure}
    \centering
    \includegraphics[width=0.8\textwidth]{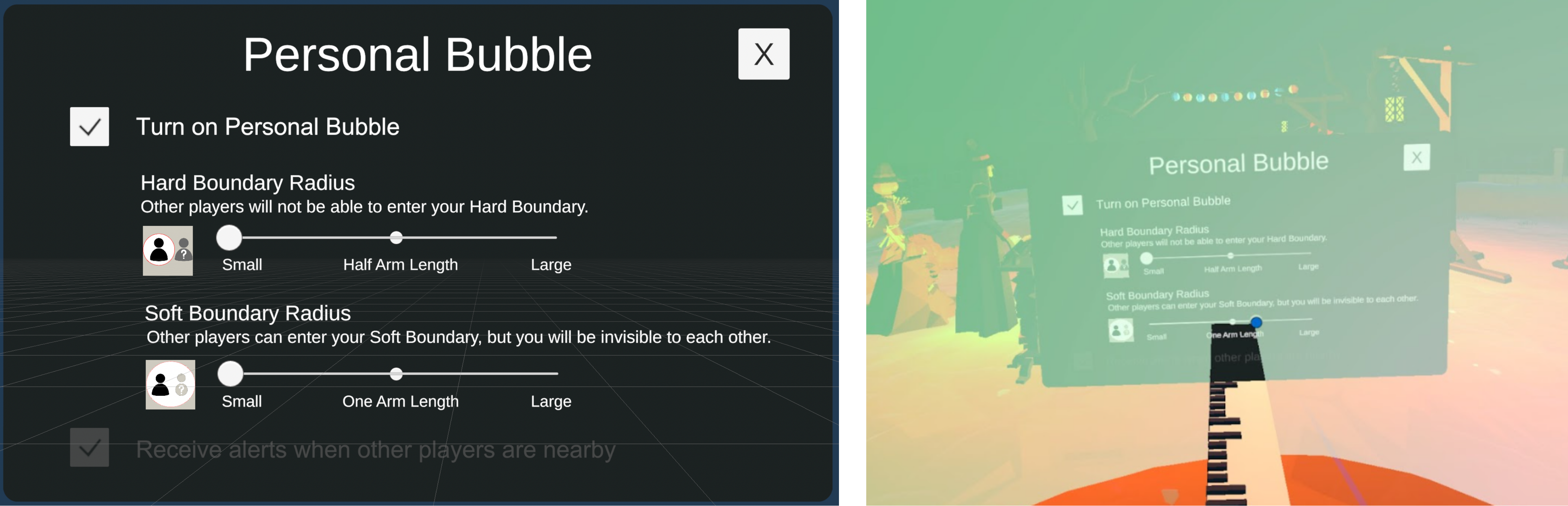}
    \caption{\textbf{Left}: Personal Bubble Menu. This menu includes the ability to select a hard or soft bubble boundary. The bubble size sliders are labeled with reference to arm-length. \textbf{Right}:  When changing the bubble size with the slider, a translucent bubble and a ruler pop up to help players gauge the distance.} 
    \label{fig:pb} 

    \includegraphics[width=0.8\textwidth]{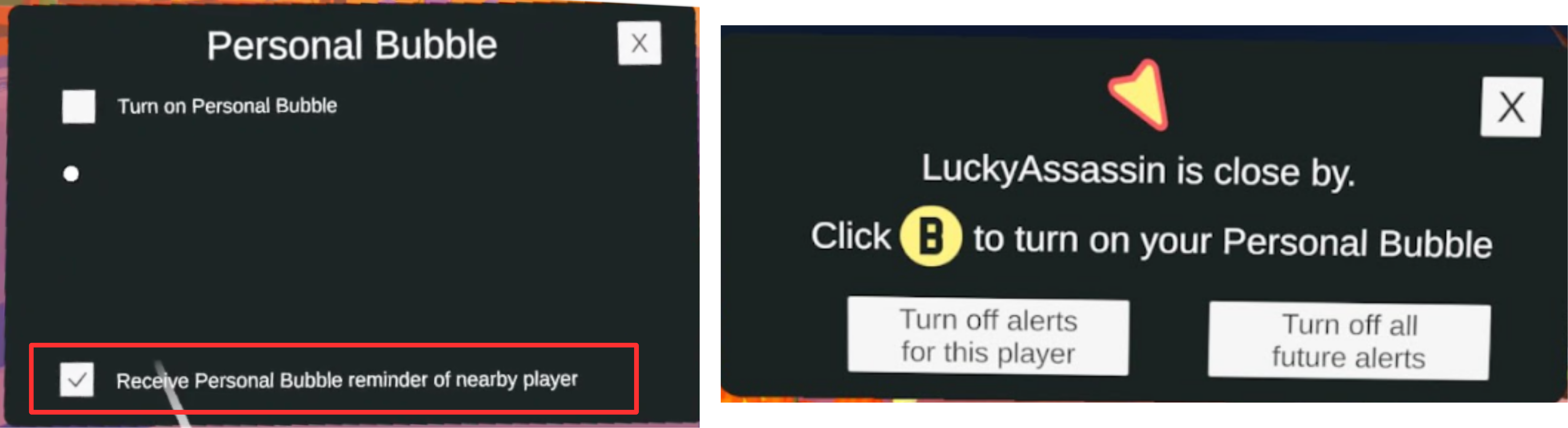}
    \caption{\textbf{Left}: Instead of the personal bubble, users can enable the ability to receive an alert when someone enters their personal space. \textbf{Right}: Alert Message when someone is near. The arrow indicates the location of the approaching player; the player can then: turn on their personal bubble with the shortcut button B, turn off alerts for the approaching player, or turn off all future alerts.} 
    \label{fig:alert} 
\end{figure}


\subsubsection{Personal Bubble}
We sought to improve the personal bubble, a popular safety feature in almost every social VR game, to make it more customizable, easier to understand, and easier to access. 
We kept the most common effect of the Personal Bubble, stopping others from entering the bubble, and call this a \textit{hard boundary}. We also incorporated the idea of an ``ignore bubble'' that exists in some social VR games, such as RecRoom, which makes both players invisible to each other when one enters another's bubble; we call this a \textit{soft boundary}. By providing both effects, players have more flexibility in deciding how to guard their personal space. 
In response to participants' confusion about bubble size, we added a translucent bubble and a measuring ruler to support gauging distance in VR, as Figure \ref{fig:pb} shows (\textbf{DG1}). We describe the size in terms of ``arm's length'' to support understanding of distance in VR. 

Finally, to design the Personal Bubble to be more context-aware and easy to access during an emergency, we added a detection feature whereby the system alerts the player when other players are close by and provides a temporary shortcut to activate the default bubble (\textbf{DG2, DG3}). In Figure \ref{fig:alert} (left), we show how the detection functionality is enabled only when the personal bubble is turned off, and  (right) we show an example of the alert message when the system detects a player close by. Players can customize the list of players they would like to avoid receiving alerts about, such as their friends.

\subsubsection{Preference Badges}
\begin{figure}
    \centering
    \includegraphics[width=0.8\textwidth]{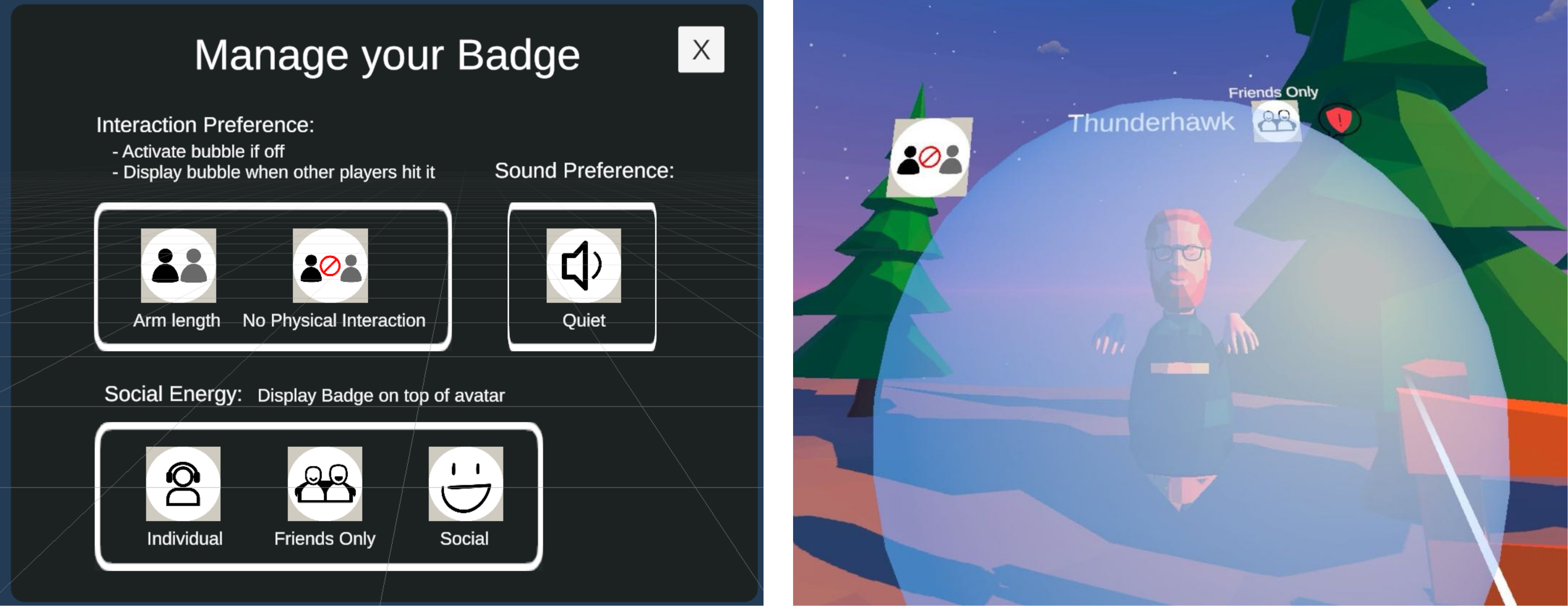}
    \caption{\textbf{Left}: Preference Badge Menu. This menu includes badge management for interactions, sounds, and social energy. \textbf{Right}: When a player selects an interaction preference badge, the badge pops up with a translucent personal bubble to communicate their preference to others. Selecting a social energy badge causes the preference to float atop the avatar's head.} 
    \label{fig:badge} 
\end{figure}

We designed a preference badge system to communicate players' preferences about interaction, sound, and social energy (Figure~\ref{fig:badge}). 

\textbf{Interaction Preference Badge}. Players signal the types of interactions they are comfortable with based on distance. `Arm-Length Interaction' means others must maintain one arm-length distance from them and they only accept interactions like handshaking and fist bumping. `No Physical Interaction' means they want no physical contact with others. To ensure preferences communicated via badges are respected (\textbf{DG6}), each interaction preference is coupled with a Personal Bubble. By activating the arm-length or no physical interaction badge, the hard boundary of Personal Bubble will be turned on to the size of an arm-length or the maximum radius, respectively. 
If another player violates this interaction preference, the personal bubble flashes for a few seconds alongside the badge to communicate that preference to others, as in Figure \ref{fig:badge}.

\textbf{Sound Preference Badge}. Players signal if they prefer a quiet environment. Like the interaction badge, they can couple this badge with audio detection of the ambient noise level that automatically reduces sound. However, since this study focuses on embodied harassment rather than noise-based harassment, we leave automated audio detection for future work. 
Thus, a sound preference badge signals only the preference and is used in the Room Preference Filter.

\begin{figure}
    \centering
    \includegraphics[width=0.7\textwidth]{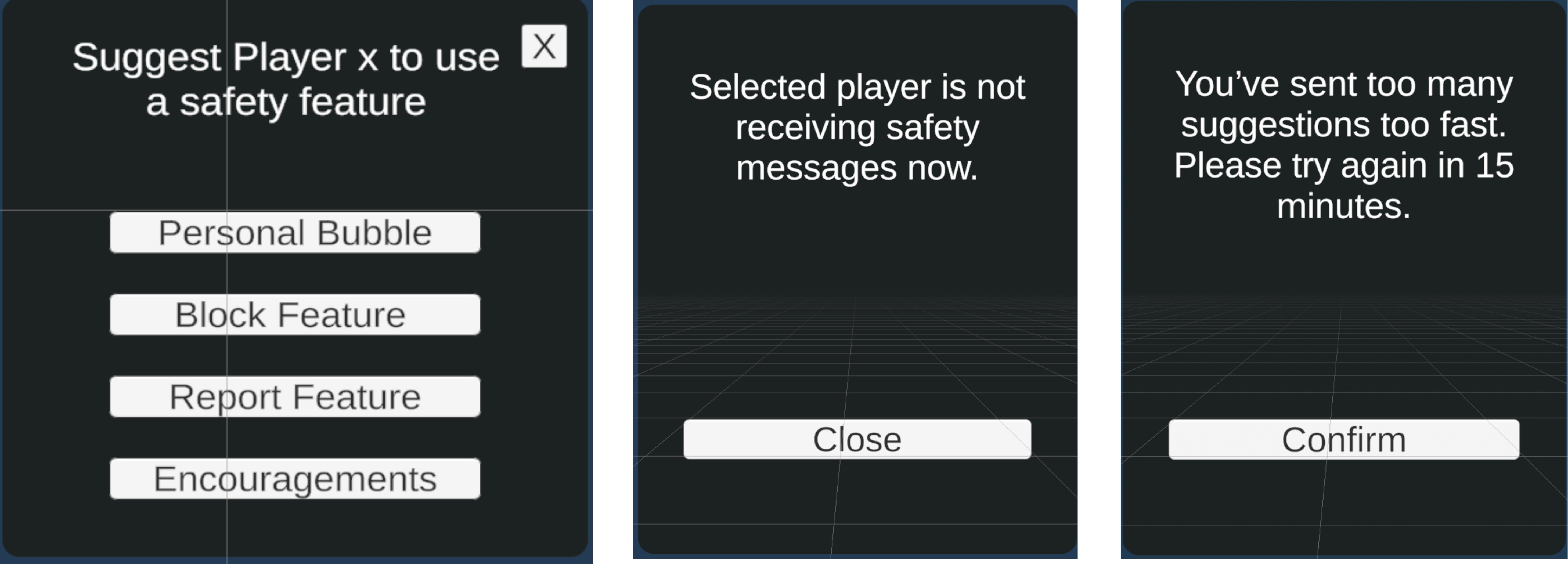}
    \caption{\textbf{Left}: Safety Suggestion Menu from the sender's perspective. The sender can suggest a safety feature to the receiver. \textbf{Middle}: The block interface when the receiver blocks future suggestions. \textbf{Right}: When system detects repeated sender suggestions, it puts the sender in a cool-down time out.} 
    \label{fig:send} 

    \includegraphics[width=0.6\textwidth]{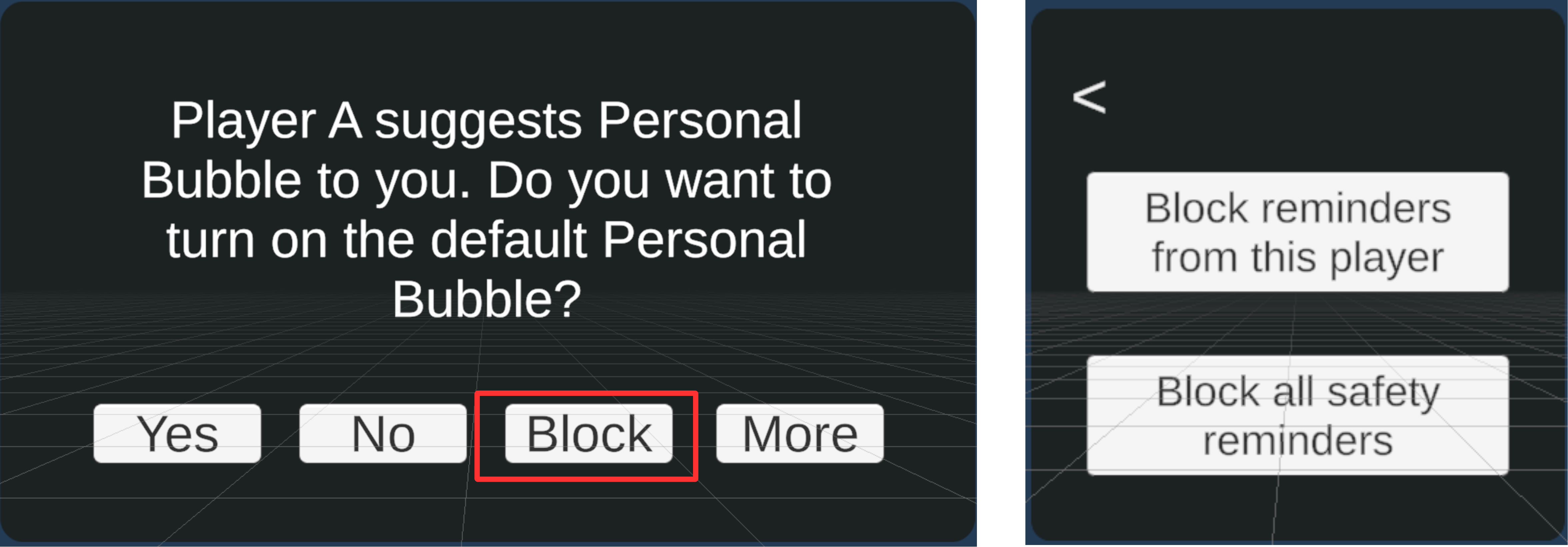}
    \caption{\textbf{Left}: Interface when player receives a suggestion from player A to use the personal bubble. Clicking ``Yes'' turns on the personal bubble. Clicking ``More'' opens the personal bubble interface for more user customization. \textbf{Right}: Clicking ``Block'' offers the option to either block suggestions from the current sender or block all future suggestions.} 
    \label{fig:receive} 
\end{figure}

\textbf{Social Energy Badge}. 
Options for signaling social energy include: `Social,' meaning the player is willing to socialize and make new friends;  `Friends-only,' meaning the player wants to socialize only with friends, not strangers; and `Individual,' meaning the player is not in the mood to socialize. Participants' diverse social needs in social VR inspired this category of badge (\textbf{DG4}), as well as similar offline tools built for the neurodivergent community \cite{Gaming_with_Hui}.
\subsubsection{Safety Suggestions}
Safety suggestions leverage bystanders to intervene during in-the-moment harassment and encourage prosocial behavior by enabling players to send each other suggestions for adopting a safety feature. They also serve as a shortcut to navigate the loaded menus in social VR games, making safety features more useful during emergencies (\textbf{DG3}).  Finally, we built safety suggestions with anti-abuse measures including limits on sending suggestions and the option to block incoming ones (\textbf{DG6}). 

From the sender's perspective, when one player is seemingly being harassed or cannot find a desired safety feature, the sender can: (1) open a suggestion menu by clicking the safety suggestions icon above the receiver's head, (2)  select the safety feature they want to suggest. The system tells the sender if the suggestion has been sent successfully or if the receiver is currently not receiving suggestions, which means the receiver is blocking either that sender or all safety suggestions (Figure \ref{fig:send} (middle)). Additionally, if the system detects the sender has sent too many suggestions quickly and may be spamming other players, the sender will be prohibited from sending for a while (Figure \ref{fig:send} (right)). 

From the receiver's perspective, a message window pops up at the bottom of the screen (Figure \ref{fig:receive} (left)) that describes the suggested safety feature and lets the receiver decide whether to accept or decline the suggestion  (\textbf{D7}).
If they are unsure about the safety feature or want more customization, they can click ``More,'' which opens the menu where the suggested safety feature is located. Additionally, if the receiver suspects ill-intention or no longer wants suggestions, they can click ``Block,'' to stop suggestions from that sender or all senders (Figure \ref{fig:receive} (right)).


\subsubsection{Room Preference Filter}
\begin{figure}
    \centering
    \includegraphics[width=0.8\textwidth]{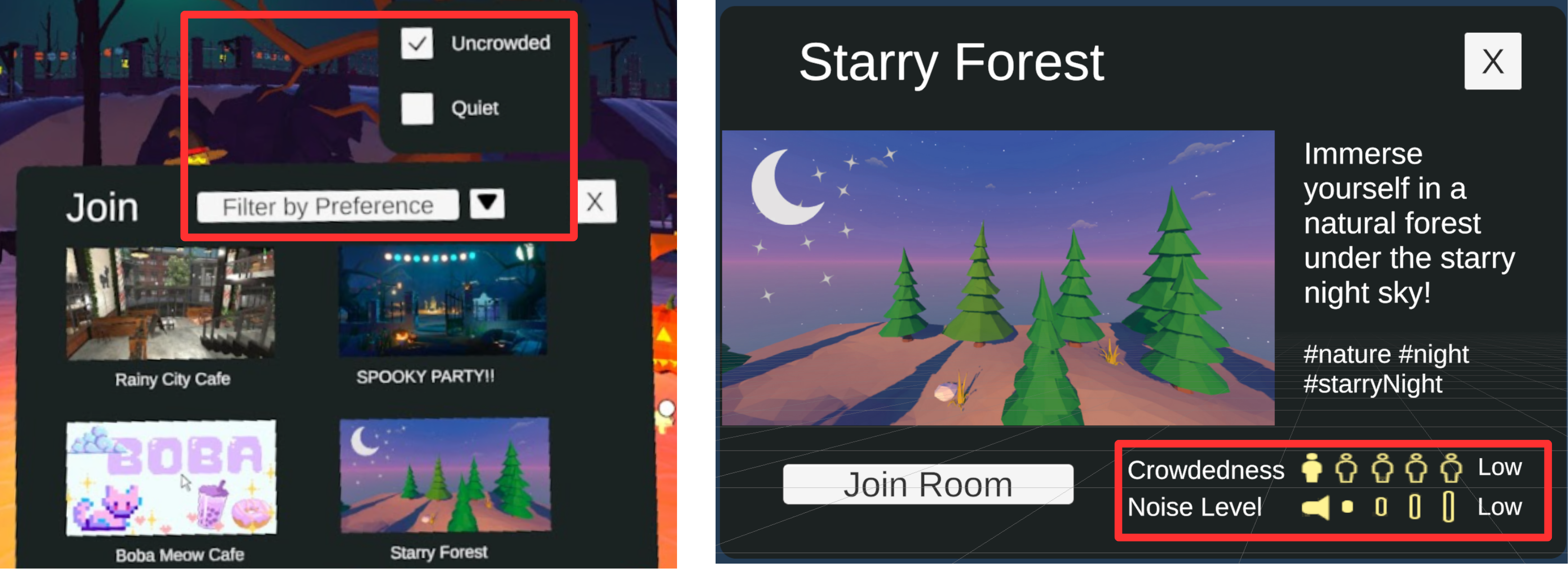}
    \caption{\textbf{Left}: Join Room Menu. By clicking ``Filter by Preference,'' players can filter for uncrowded and/or quiet game rooms. The filter preference is automatically set to match the interaction and sound preference badge if selected. \textbf{Right}: Interface for each game room. Icons on the bottom right indicate the current crowd and noise levels in the room.} 
    \label{fig:room} 
\end{figure}

This feature helps players find a game room that meets their preferences for crowd and noise levels (\textbf{D5}). 
If the player displays any preference badge, the filter automatically aligns with those preferences to help them find a desired room (Figure \ref{fig:room} (left)). By clicking on each room's information menu, icons on the bottom right indicate crowd and noise levels in this room, which should be updated based on the number of gamers in the room (Figure \ref{fig:room} (right)). 

\subsection{System Implementation} 
The system was implemented using the Unity Game Engine, the industry standard for VR development. We chose to develop the game for Oculus Quest2 due to its popularity among VR users and the availability of testing headsets. We adopted the Meta XR All-in-One SDK package for fundamentals and the Photon Unity Networking for multi-player connectivity, which supported users joining a virtual server and synchronized player changes such as selecting a new badge.
To implement the game rooms, we used components from BOXOPHOBIC’s Skybox Extended Shader package to build a starry forest scene and components from Polyperfect to build a Halloween-inspired scene. Users can transition between rooms through teleporting. We built player avatars with the City People Lite asset created by Denys Almaral, simulating a more authentic social VR gaming experience.

%% file: 5_evaluation.tex
\section{Evaluation}

To evaluate Puffer, we conducted a user study to test usability and evaluate whether the system meets the design goals. We also aimed to identify how the system impacts players' responses to \textcolor{edits}{embodied} harassment and the social dynamics in social VR games. We recruited diverse participants with various gaming experiences, primarily focusing on demographics more susceptible to harassment. We sought to answer the following questions:
\label{evaluation_scale}
\begin{description}
    \item[RQ1: Prosocial Norms.] How does the system cultivate positive prosocial norms in the game to proactively dissuade harassment from happening?
    \item[RQ2: Effective Safety Protection.] How effective is the system in protecting players in (1) the context of when an emergency arises (e.g., players are actively being harassed),  and (2) the general context of playing the game?
     \item[RQ3: System Robustness Against Misuse.] How robust is the system when malicious players aim to abuse its functionality?
    \item[RQ4: Usability and Learnability.] How usable and learnable is the system?
\end{description}

\subsection{Participant Recruitment and Demographics}

We recruited 16 participants by posting in online gaming and VR communities on Reddit (r/VRGaming, r/oculus, r/OculusQuest, r/VRChat, r/GorillaTag, r/RecRoom, r/HorizonWorlds, r/transgamers) and student groups at a major US-based university, and then used snowball sampling (Table \ref{tab:demographics}). P3 and P4 previously participated in the formative study and were invited again to the user study. 
\footnote{\textcolor{edits}{P3 and P4 were women gamers with a zero-trust mindset and expressed concerns about the misuse of safety features by malicious players in the formative study. By reinviting them, they continued to evaluate Puffer, especially the anti-abuse mechanisms we added based on their feedback.} }
To reach participants who are more likely to be impacted by harassment in gaming, we connected with affinity groups at our university, such as Women in Computing, Society of Women Engineers, and a group for LGBTQ+ computer science students.
As Table \ref{tab:demographics} shows, 68.75\% of our participants identify as women or LGBTQ+, demographic groups more likely to encounter sexual and identity-based harassment  \cite{Freeman2022, womenVR}.  

\input{Participant_table}

\subsection{Study Procedure}
Each user study session spanned about 90 minutes and was conducted in person at a major US university campus.
\footnote{\textcolor{medits}{Because Puffer requires an Oculus Quest or Quest2 headset to operate, we could only accept participants who could join in-person session or owned a Quest headset for a remote session. We had to turn down one participant who did not satisfy the requirement.}}
We invited two participants for each session to simulate social interactions in a social VR game. A total of eight sessions were conducted. Participants could invite friends to conduct the session, and participants in three sessions were friends, and the rest were randomly paired. Two researchers were present to guide the session, help participants wear the Quest2 headset, and take observation and interview notes. \textcolor{edits}{Each session included an interface exploration section, two simulated scenarios, and a debrief interview with an exit survey.}

\textbf{Interface Exploration}. First, participants were asked to \textcolor{edits}{freely explore the Puffer interface without receiving instructions or assistance}. 
\textcolor{medits}{At the end of the exploration}, we asked players to complete four tasks to test learnability. These tasks simulated player motivations and safety needs in a social VR context. Participants selected Puffer features to address those needs, \textcolor{medits}{where researchers evaluated their understandings of the features.} (see Appendix \ref{apx:interface_exploration} for more details). 
After completing the tasks, participants watched a tutorial video (a similar video can be found in the supplementary materials) formally introducing all features and were given time to ask researchers feature-related questions to clear any confusion before moving on to the simulated scenarios later.

\textbf{Simulated Scenarios}.
Participants next engaged in \textcolor{edits}{two scenarios that simulated real gaming situations where players would adopt safety features.}

\begin{itemize}
    \item \textcolor{edits}{\textbf{Roleplay Scenario}. This scenario tested Puffer in varied social contexts. Two participants were randomly assigned to roles with opposite interaction and social preferences and were unaware of each other's preferences. They conducted three tasks: (1) trying to make friends with each other, (2) finding game rooms that matched their role's preferences, and (3) acting as bystanders to observe interactions between two researchers (additional details in Appendix \ref{apx:roleplay})}.

    \item \textcolor{edits}{\textbf{Tagging Game Scenario}. This scenario tested Puffer under \textcolor{medits}{time-sensitive conditions}. Researchers acted as taggers, and participants needed to avoid being tagged by activating their personal bubble. They were given three ways to access their bubble and assessed the convenience of each (additional details in Appendix \ref{apx:tag}).}
\end{itemize}

\textbf{Debrief Interview and Survey}.
After the scenarios, participants shared their in-depth feedback about the system through a semi-structured debrief interview. \textcolor{edits}{They discussed each feature separately with the same set of interview questions and were encouraged to compare features. The interview questions focused on examining the usage scenarios for these features and their potential positive or negative impacts on players and the broader gaming environment.}  For preference badges and safety suggestions, we also asked additional questions about the two perspectives for these features (e.g., wearing vs seeing a badge and sending vs receiving a suggestion) to capture nuances about how safety features affect player relationships. 

Participants also answered survey questions to provide quantitative feedback about each safety feature. 
The first 10 questions were derived from the System Usability Scale (SUS), a widely recognized questionnaire for measuring perceptions of usability \cite{sus}. \textcolor{edits}{The remaining questions collected participants' opinions on feature effectiveness and impact. 
Both the interview and survey questions were designed in response to the evaluation questions to better assess Puffer from the scales in \ref{evaluation_scale}.}
The complete interview and survey protocols are in Appendix \ref{apx:questions}. 

\textcolor{edits}{ \subsection{Ethical Considerations} Our study was reviewed and exempted by the IRB of a major U.S. university. Since the study involved sensitive subjects like harassment, we followed a trauma-informed approach to user research \cite{trauma-chen}. Acknowledging the possibility of (re)traumatizing participants who previously witnessed or experienced harassment in games, we avoided placing them in simulated harassment scenarios, instead designing role-playing and game-like scenarios. We also chose to test in a controlled environment instead of actual social VR games to maintain control over what would be exposed to participants. Further, participants could choose in-person or virtual settings for the study and were encouraged to bring friends to enhance their feelings of safety and familiarity. Finally, adhering to the collaboration principle in trauma-informed computing, we sought to involve players who previously experienced harassment in social VR but did not request them to share those experiences \cite{trauma-chen}.}

\subsection{Data Analysis}
 We conducted \textcolor{edits}{an inductive} thematic analysis to interpret the observational and interview data \textcolor{edits}{\cite{thematic_analysis, reflexive_thematic_analysis}}. Two researchers collected observational data, including participants' behaviors and reactions during the interface exploration and simulated scenarios. 
 \textcolor{edits}{
 After every session, two researchers compared their observation and interview notes, ensuring consistency in documented participant behaviors, and discussed notable reactions or responses. The first author recorded, transcribed, and manually corrected qualitative interview data. She then closely read all transcripts and observation notes, marked sections of interest, and derived a set of initial codes. These codes were applied to the marked sections and categorized to generate themes, which were then shared with the last author. Through discussions, authors regrouped codes and refined themes, producing a final codebook of 36 codes that captured feedback on safety feature designs and the potential impact of these features on players and the gaming environment. Example codes include ``Decrease crowdedness/visual noise,'' ``Encourage norms of looking out for each other,'' and ``Increase stress or overwhelmingness during emergencies.'' The complete codebook is in Appendix~\ref{final_codebook}}. 

%% file: Participant_table.tex
\begin{table}
\centering
\small
\begin{tabular}{p{0.3cm} p{0.3cm} p{1.5cm} p{1cm} p{2cm} p{3.6cm} p{3.6cm}}
\toprule 
\textbf{Pid} & \textbf{Age} & \textbf{Gender} & \textbf{Race} & \textbf{Part of LGBTQ+ Community} & \textbf{Social VR Gaming Experience} & \textbf{Other Gaming Experience} \\
\midrule
3  & 23 & Woman & Asian, White & Queer & N/A & Superhot \\ \hline
4  & 23 & Woman & Asian & N/A & Rec Room, Echo VR, Gorilla Tag & Beat Saber, Superhot, Apex Construct \\ \hline
9  & 24 & Woman & Asian & Bisexual, Transgender & VRChat (100+ hours) & N/A \\ \hline
10 & 19 & Man, Woman, Non-binary & Asian & Bisexual, Transgender, Queer, Asexual/aromantic/agender & VRChat (20+ hrs), Horizon World (2 hrs), Rec Room (2hrs) & Phasmophobia (4-8 hrs) \\ \hline
11 & 26 & Man & White & N/A & Rec Room (6 hrs), Gorilla Tag (1 hr) & Cook-out (3 hrs)\\ \hline
12 & 22 & Woman, Non-binary & Asian, White & Lesbian & VRChat (3 hrs) & Beat Saber (10 hrs) \\ \hline
13 & 19 & Woman & Asian & Bisexual & Shooting game cannot recall name (6 hrs) & N/A \\ \hline
14 & 21 & Man & White & N/A & N/A & Beat Saber, Gorn, League of Legends, Overwatch \\ \hline
15 & 21 & Non-binary & White & Asexual/aromantic/ agender & VRChat (7 hrs) & Phasmophobia, League of Legends, Overwatch \\ \hline
16 & 22 & Non-binary & Hispanic or Latino, White & Bisexual, Transgender, Queer & N/A & Beat Saber, Half-life: Alyx \\ \hline
17 & 19 & Man & White & N/A & VRChat (1450 hrs) & N/A \\ \hline
18 & 20 & Woman & Asian & N/A & N/A & Plato (20 hrs), Genshin (100 hrs), Fall guys (50 hrs) \\ \hline
19 & 18 & Woman & Asian & N/A & N/A & League of Legends (3 hrs per day) \\ \hline
20 & 19 & Woman & Asian & N/A & N/A & N/A \\ \hline 
21 & 21 & Man & Black & N/A & VRChat (20 hrs), Horizon World, Gorilla Tag & Big Screen Beta \\ \hline 
22 & 19 & Man & Asian & N/A & N/A & GTA 5 (1500), Battlefield 5 (300), Call of Duty (200) \\
\bottomrule
\end{tabular}
\caption{Evaluation Study Participant Demographics and Gaming Experiences}
\label{tab:demographics}
\end{table}


%% file: 6_results.tex
\subsection{Findings}

\subsubsection{\textbf{RQ1: Prosocial Norms.}} \textcolor{edits}{Participants identified opportunities for Puffer to improve unclear social norms and avoid potential conflicts, creating a more friendly gaming environment for positive interactions.  }

\textbf{Participants found that sharing preference information helps clarify ambiguous social norms during social interaction.}
In the roleplay scenario, participants who acted as a friendly experienced player (named Thunderhawk) were instructed to make friends with participants who acted as a nervous new player (named Peppermint), \textcolor{edits}{who wore `Individual' and `No physical interaction' badges}. Thunderhawk participants exhibited three behaviors.
\begin{enumerate}
    \item \textcolor{edits}{\textbf{Attempted to interact.}} Some Thunderhawk participants tried to initiate a conversation to further understand Peppermint's preference and gauge if they would like to talk. When Peppermint responded in a friendly way, for example, counter to the `Individual' badge they wore, P21 realized that P22 lacked previous gaming experience or knowledge about safety features. P21 tried to provide gaming suggestions and teach P22 to use some features.

    \item \textcolor{edits}{\textbf{Understood the rejection.}} Some Thunderhawk participants approached Peppermint and were stopped by the personal bubble and rejected by Peppermint for interactions. The Thunderhawk participants held no negative emotions after rejection because Peppermint's preference was shown via the badge. As P19 shared, ``\textit{It was clear to me that Peppermint just wants to be on their own...I did not feel offended when rejected.}'' 

    \item \textcolor{edits}{\textbf{Respected the preference.}} Some Thunderhawk participants fully respected the Individual badge and decided not to have any interactions, such as P16. They shared that they would try to make friends with other players in the room who had a social badge instead.
\end{enumerate}

In general, preference badges mediated the social interactions between players with opposite preferences, preventing unintentional conflicts caused by unclear preferences. \textcolor{medits}{Thunderhawk participants perceived the badge both proactively, by not interacting with Peppermint, and reactively, by accepting the rejection easily. Through these interactions, badges helped normalize boundary setting between players.}

When debriefing, participants shared that preference badges would be useful in social VR games where players employ various personal strategies to improve their experiences that are unknown to others. With badges signaling their preference, players with diverse preferences and needs can coexist respectfully with fewer misunderstandings. 
For example, P4 discussed how a social energy badge could help him avoid conflicts: ``\textit{[The badge] could protect against future harassment, too, if somebody thinks that you are rude because you are not replying [due to turning off the audio] and when you come back to turn on the sound and suddenly there are people harassing you.}''
Also, participants echoed that requiring everyone to verbally negotiate personal preferences can take effort and escalate into conflicts when misunderstandings happen. P3 highlighted the ease of communicating preferences through badges:
``\textit{The unspoken understanding between players without having to explain verbally what your preferences are, and just being able to see it from a distance, is very helpful. 
I think it helps set hard boundaries...without the option to negotiate them, which I think is really important.}''




\textbf{Participants felt that proactive features would help players find other players and rooms that match their preferences.}
Multiple participants discussed how the social energy badge supports a smoother, easier social experience with other unknown players, streamlining the process of finding similar-minded players: 
``\textit{They definitely help social people find social people and not unintentionally harass anti-social people.''} (P16)
P21 highlighted that the social energy badge also supports new players \textcolor{edits}{in making friends and} blending into the new community more quickly: \textit{``\textit{[the badge] will be great for new people who are trying to get to VRChat, those are the alerts that they will look out for communication,…to balance the dynamics and understand who to talk to.}''}

Similarly, P10 shared that the room preference filter helps players find game rooms with a particular vibe, leading to a more personalized gaming experience. ``\textit{I think [this feature] has a positive impact by letting people easily find the kind of vibe they are looking for instead of just going to the most popular ones. I think that happens in VRChat...a lot of people just going to the top 5 most populated rooms no matter what they are looking for, crowdedness or quietness.}''
P4 praised the transparency of a room environment given the room preference filter: ``\textit{...it gives the same `you know what the vibe of the room is going to be like'...You can see if it’s crowded or quiet before you go into it, which is quite nice.}'' 
By integrating these functions into the system, players are supported with transparent expectations when engaging with others and in the game rooms: ``\textit{The badge would be more preventative…setting the expectations in the space,}'' as P16 observed. 


\textbf{Participants felt that the presence of safety suggestions can encourage mutual support.}
Participants noted that features like safety suggestions
also facilitate positive social interactions and norms. 
P14 mentioned that even if the help received may not be necessary, it signals thoughtfulness on the part of the bystander and sparks positive conversations: 
``\textit{After you've done whatever you needed to do to the person that was annoying you, you can turn to this other person and go, `Oh, thank you for doing that. That was actually pretty helpful.' That sparks a conversation that is actually well-meaning.}'' P10 spoke from the perspective of new players: 
``\textit{If you are a new player and some people use this feature to help you out, that could help them feel better, that other people are on your side. There are people looking out for me…helping me to make me safe...that can feel good.}'' Such support not only diminishes feelings of isolation and loneliness but could also cultivate a sense of friendship and camaraderie in the community.

In addition to promoting positive social interactions, safety suggestions could encourage players to be more observant of other players. 
Without system support for intervening, bystanders who intervene could themselves become targets for retaliation by offenders.
As P14 pointed out, safety suggestions avoid escalating the situation, where offenders could become aggressive towards the bystander, too, which he had observed in previous gaming experiences. 
By reducing barriers to intervention, players can more easily act as friendly bystanders, \textcolor{medits}{fostering a greater sense of responsibility} for helping each other. As P3 noted: ``\textit{...more players looking out for each other, and maybe being more aware of less favorable things happening around them [and] being more observant to what other players are going through.}'' \textcolor{medits}{Safety suggestions could empower players to support each other, rather than relying solely on moderators.}



\subsubsection{\textbf{RQ2: Effective Safety Protection.}} \textcolor{edits}{Participants found Puffer simplified ways to protect themselves or help others especially under urgent situations.}

\textbf{Participants found instant-reactive safety features quick and easy to access under time pressure.}
During the tagging game scenario, participants tried three different ways to turn on the personal bubble: (1) the menu, (2) the shortcut provided by the alert message, and (3) the safety suggestion. In general, more participants turned on the personal bubble before being tagged with the shortcut provided by the alert message than by navigating through the menu.


The hotkey provided by the alert message was easier to press than menu buttons, and some participants suggested having permanent hotkeys for safety features to increase their ease of access. In contrast, without the alert message, \textcolor{edits}{participants found it extremely hard to negotiate multiple clicks to access the bubble via menus, especially in the moment of harassment,} as P12 shared, ``\textit{Opening the menu is only doable when being away from [the] tagger and not in a rush}.'' Another benefit of the alert message was the arrow, which assisted players in locating the tagger and potential danger. Without the message, players were unaware when tagged from behind. 

For the safety suggestions approach,
participants who received a suggestion from another player found it quick and easy to accept. 
During the debrief session, they reflected on how safety suggestions from friends could help players who are unfamiliar with safety features and menu systems. P16 shared, 
\textit{``It's like they [their friend] can send the reminder, and you can just hit one button and it turns on the bubble. You don't have to worry about digging through the menus in order to find your bubble.''}
Participants with experience using other online multiplayer games commented on how the private nature of safety suggestions helps with de-escalation, preventing players from arguing in a public chat. Per P15, an experienced Overwatch and League of Legends player:  ``
\textit{Instead of somebody saying just mute them, if they send me the message privately and I would just be able to mute them from there, that would definitely be a good feature.
…I play [League of Legends], [where] people would be flaming in the chat all the time; people would be arguing. If they are typing, they are not playing, so they are distracted from the game itself…}''


\textbf{Preference badges helped participants understand social context and created opportunities for bystander intervention.}
During our formative study, participants expressed concerns about bystanders misreading the social context and intervening inappropriately. Due to the lack of consensus about norms and the blurry boundary between fun/play and harassment, bystanders have difficulty identifying harassment and consequently lose opportunities for support due to hesitation. 
During roleplay, when participants acted as bystanders, the researcher's `No Physical Interaction' badge affected participants' reactions to the scenario. When badges were provided, six of eight participants identified harassment; otherwise, three of eight identified harassment. 
When debriefing, participants shared how badges impact their interpretation of the scenario. 
As P12 shared, ``\textit{{The combo of the bubble popping up over and over and also the badges [made it] very clear that [the researcher] didn’t wanna interact.}}'' This effect is shown in Fig \ref{fig:badge} (right). 
P12 also mentioned how the badge changed their reading of the context, ``\textit{That sign was more obvious, more indicating. At first, I thought maybe they are just playing the games…Having the badges made it clear that it wasn’t two friends.}'' 
When badges were not present, participants needed more verbal cues to decide whether harassment was occurring. P22 shared that harassment ``\textit{did not come into my mind,}'' and P16 shared, ``\textit{[the researchers were] maybe just hanging out with each other, without a badge; maybe it’s not my place to step in. I could be annoying to somebody.}''


Similarly, during the debrief interview, participants noted that the proactive features were not limited to communicating preferences but were effective in supporting bystanders in intervening. As P16 said, ``\textit{Not only do [badges] help you figure out if somebody wants your interaction, but it helps you figure out if somebody wants somebody else’s interaction. In the scenario where you two were harassing each other, I didn't know if you guys were friends. If one of you had an individual badge, I would be like, hey, use a bubble.}''

\subsubsection{\textbf{RQ3: System Robustness Against Misuse.}}

While participants acknowledged the anti-abuse mechanisms built into the system, they also mentioned more opportunities to accommodate the complex dynamics in social VR gaming (see survey ratings in Appendix~\ref{apx:ratings}). 


\textcolor{edits}{\textbf{Misinterpretations of safety suggestions due to lack of context.} } To avoid malicious usage, we chose to design safety suggestions with only pre-programmed text as opposed to allowing participants to edit messages. \textcolor{edits}{Because of the pre-programmed nature, the suggestions can be seen as lacking specific contextual information, such as the rationale for sending the suggestion, potentially leading to misinterpretation if the suggestion is mismatched with the receiver's expectation.}
Such mismatches could feel awkward for experienced players who already understand the safety features: ``\textit{If someone is harassing me and I get a suggestion to block this player, it could be helpful for people who don’t know this [block] functionality, but it could also feel a little bit like passive aggressive…? If I were an experienced player, I would just turn that off in my settings}'' (P10). 
P19 shared that if the sender's intention is unclear, an irrelevant suggestion message could create doubt: ``\textit{If you are in the middle of playing and someone tries to suggest [for] you to use the personal bubble, you might think that something is going wrong. 
If the context is weird [when I receive the suggestion], I would suspect it’s trolling. That would create doubt among players. It’s very useful if the context is right.}''



\textcolor{edits}{\textbf{Some safety features can lead to a reduced user experience quality and increased stress.} The hard boundary effect of personal bubbles is not always feasible because physical interactions are essential to the VR game experience. ``\textit{I think a lot of VR games do require some level of physical interactions, such as the sporty ones}'' (P4). P10 also shared that they enjoy giving and receiving ``head pats" from other players in VRChat, a common interaction to show friendliness. }

Instant-reactive features could also increase stress or overwhelm users.
Participants noted that players experiencing harassment could be under significant stress and unable to react to an extra message or window pop-up if they are not familiar with these messages. 
P12 commented on safety suggestions: ``\textit{I would worry that potentially, it's like unhelpful under, you know, could add extra stress. What if they already do know [about] that feature? If they're like, overwhelmed already with [the situation].}'' 
P4 mentioned that the alert message from the personal bubble could also be overwhelming: ``\textit{It was kind of overwhelming to have that menu [alert message] pop up right then [and to have] to make the decision whether or not I wanted to turn on the bubble. I was like, `Which one do I choose here? I'm not sure.' I cannot tell if he [the stranger] is friendly yet…So that was a little bit overwhelming at the very beginning, especially as he was approaching me. I don't have enough time to decide.}'' \textcolor{edits}{P4 also commented that the alert messages could be distracting if they pop up too often in a crowded room, causing players to turn them off and ``regret that later."}  


\textcolor{medits}{\textbf{
Some safety features turn players' personal preferences into targets for abuse, making marginalized players susceptible to harassment.} }
Participants highlighted that preference sharing could lead to unwanted attention and even danger. For example, P12 shared from the standpoint of being a woman gamer, \textit{``If I am presenting as a woman [in my avatar], I would absolutely [not use the social badge]…and I would be worried that being a targeting thing, saying I am `Individual.' }'' 
\textcolor{medits}{Women participants noted sharing preferences might conflict with their strategies to protect or disguise their information online to 'avoid trouble', highlighting that the cost of sharing varies for players with different identities.}
Instead, they suggested swapping the word `Individual' in the social energy badge with other terms that are less likely to place them in a disadvantaged situation, such as `low energy,' `busy,' or `absent.' 
Additionally, they noted that connecting badges with safety features is a helpful strategy. 
The `Individual' social energy badge could be bundled with features such as mute and personal bubble, or it could give players the ability to customize the safety features connected with it. 

\subsubsection{\textbf{RQ4: Usability and Learnability.}}
Puffer was straightforward to learn without assistance for most participants, especially those with prior VR experience. 
\textcolor{edits}{Thirteen of sixteen participants understood its features and adopted the correct features in all tasks.}
Three participants, P18, 19, and 20, misunderstood some safety suggestions and the difference between hard and soft boundaries, or they failed to notice the room preference filter. These participants had zero VR gaming experience before the session and were able to clarify their confusion after watching a tutorial video. 
From the SUS scale, the preference badge and room preference filter show high usability, but safety suggestions need improvement (see ratings in Appendix~\ref{apx:ratings}). This aligns with participants' feedback, where most of Puffer was straightforward except for safety suggestions, which they had never experienced in other games, were confusing without a proper introduction. Based on participant feedback, we iterated on the safety suggestions feature during our evaluation study, and the detailed process is included in Appendix~\ref{apx:iteration}.


%% file: 7_discussion.tex
\section{Discussion}
Our findings demonstrate that proactive and instant-reactive designs can help players address issues like harassment in social VR in ways that are difficult to address via solely reactive approaches.
We discuss how these features not only help to reduce harm by preventing and stopping harassment in the moment but, by clarifying norms and helping players find other compatible players, they also broadly support having more positive interactions.
 We then discuss further considerations when designing safety features that take into account marginalized users' experiences and existing strategies and argue for designing safety features that are voluntary and maximally flexible.
 


\subsection{Beyond reducing harm: promoting positive and enjoyable social experiences}

A significant challenge when mitigating social VR harassment, argued by prior work, is the ambiguity of social norms, where players are unclear about expected behaviors from others and from the platform \cite{blackwell2017, Freeman2022}. Causes of this ambiguity include the diversity of gaming environments and the lack of ways to assess norms. Prior research found that community guidelines were not integrated into the games, and players needed to learn the norms by observing other players' behaviors \cite{blackwell2017}. Additionally, each game presents a unique community \cite{Freeman2022}, and within even one game, norms vary dramatically depending on the room theme and built-in aesthetics, such as a dance floor versus an art gallery. Therefore, players need to adapt their expectations and behaviors according to the specific gaming environment they are entering.

Another cause of ambiguity is the lack of consensus on what constitutes harassment in VR, as seen in the debates about harassment between VR players \cite{Sparrow}. The perception of what constitutes harassment and how to react to it is personal and connected to one's identity and perceived power in the game \cite{Zheng2023}. 
Perceptions can be impacted by various factors, including the likelihood of encountering harassment, perceived harm and consequences of harassment, players' knowledge of the game, their support network and community, and their prior experiences or trauma. For example, during our formative study, women participants expressed concerns about interactions leading to escalation or retaliation, especially when intervening in harassment, whereas male participants were ambivalent. \textcolor{medits}{Our proactive features, preference badges and room preference filters, help reduce these ambiguities by clarifying acceptable interactions and expected norms in the game rooms. \textit{Priming} techniques, such as requiring players accept terms and conditions, were found to enhance self-moderation and build norm awareness \cite{tametoxicplayer}. Similarly, these proactive features act as embedded primes for community guidelines and personal boundaries and are more actionable and prominent in the system. }
Our results suggest that proactive and instant-reactive measures not only mitigate negative experiences but also promote positive ones.
Proactive features encourage and foster prosocial behaviors, and as more players adopt them, they could contribute to a more respectful social VR environment.
Likewise, instant-reactive features, such as safety suggestions, can create positive social experiences and build prosocial norms of mutual help. 
\textcolor{medits}{These features are particularly beneficial for new players, who are less advantaged with fewer self-protection strategies due to the lack of support systems\cite{womenVR}. For example, participants found preference badges could help them make friends more easily and safety suggestions enable experienced players to assist them during negative experiences, facilitating their integration into the social VR community. }


Participants perceived our features to improve game experiences and create more fun beyond safety protection. 
Experienced social VR gamers in our study, though less worried about harassment, still valued the preference badge and room preference filter for helping social people connect with each other. They also appreciated the social energy badge, which allowed them to signal social intentions and encourage more interactions. 
\textcolor{edits}{These insights encourage us and other designers to reflect on how safety features in games could move beyond a damage-centered design to embrace  possibilities for bringing the gaming community together \cite{flourish}.}

\subsection{Designing safety features that complement marginalized players' existing proactive and instant-reactive strategies} 
In both our formative and evaluation studies, participants who identified as women and queer compared our designs with their existing strategies and commented on the compatibility. Facing a hostile virtual environment and a lack of support from platforms, women users have long implemented their own strategies to mitigate harm and harassment across a variety of online social spaces \cite{womenVR, Vitak, coteWomenGamer}. 
This valuable knowledge is produced by experiencing the power dynamics that place them as a nondominant group, according to feminist standpoint theory \cite{standpoint}. 
Women players have been practicing both proactive and instant-reactive safety strategies \cite{womenVR, coteWomenGamer, Fox}. 
For example, proactive strategies include disguising or camouflaging their identities, avoiding strangers, and decreasing engagement. Instant-reactive strategies include not responding to confrontation and quitting the game. 
Despite some tactics contributing to negative stereotypes or sustaining a vicious cycle of toxicity in games, it is unreasonable to expect people to surrender their hard-earned strategies and knowledge. 

Thus, we maintain that proactive and instant-reactive designs should center marginalized players' experiences and not conflict with their existing strategies.
During the formative study, when evaluating bystander intervention strategies, 
some women and queer participants questioned a direct intervention approach such as ``physical'' interactions between bystanders and victims because they did not feel comfortable either stepping in or accepting a personal bubble from a stranger. This conflicted with their mindset of avoiding confrontations with strangers. However, the indirect messaging approach, which evolved into our safety suggestions feature, was considered more acceptable because it de-escalated the situation without making the offender potentially more aggressive. Although we recognize the trade-off of losing the opportunity to educate the offender because of the lack of confrontation, we prioritized not placing marginalized players in an uncomfortable and unsafe position. 


In general, we found that designing for bystander intervention was fraught with trade-offs.
Some participants who proactively avoid strangers were reluctant to accept \textit{any} help from an unknown bystander, even from our safety suggestions feature. Participants pointed out possibilities for misusing or misinterpreting a bystander system,
which has occurred in other online games, such as the endorsement system in Overwatch2 and the honor system in League of Legends. Researchers and players identified sarcastic and fraudulent endorsements, such as endorsing an enemy who played poorly to make fun of them \cite{overwatch, lolhonor}. 
One way to potentially address this would be to present distinctions between bystanders based on the players' level of trust towards them instead of putting everyone under the umbrella of bystanders. For the safety suggestions in Puffer, knowing that the source of help is a friend would increase the possibility of players accepting the help. This system also helps experienced players introduce their friends to a new game. Though limiting the feature to friends would reduce the potential for a broad set of bystanders to positively intervene, it may be an acceptable trade-off that complements people's existing strategies.

Additionally, designers should closely assess the intersectional impacts and consequences of their designs. When evaluating preference badges, P12 noted that she would feel uncomfortable wearing an ``Individual'' badge when representing herself in a feminine avatar. Despite ``Individual'' being a neutral description of one's social state, pairing it with an obvious indicator of gender could lead to unwanted attention and harassment. Designers should design with these considerations in mind given the pervasive lack of trust between players that can lead to unintended negative consequences. 


\subsection{Enabling customization of safety features and keeping usage voluntary} 
Since there is no single or ``authentic'' experience of being marginalized, safety needs can be occasionally contradictory. 
During our evaluation, we found that among all participants who identified as women or LGBTQ+, they expressed distinct or even contradictory attitudes towards safety designs due to their personality, previous gaming experiences, and more. 
For instance, some were concerned that the preference badge could lead to unwanted attention and preferred to couple it with other safety features, such as personal bubbles, while others appreciated the badge but did not want to connect it with personal bubbles because it could prevent friendly embodied interactions.
Additionally, these user needs are not fixed but dynamically change for each individual based on their experiences. Participants shared that some safety features would have been helpful when they first tried the game as a newcomer but are not as useful now since they were familiar with the game and had supportive friend networks. 

Thus, we maintain that safety features should (1) focus on supporting players in making choices that they are confident and comfortable about, and (2) provide players with multiple ways to make that choice effortlessly instead of automating the process for convenience, which could be a potential pitfall of proactive protection. This also aligns with prior research's discussion of leveraging consent mechanics and not stripping users of their agency when interpreting their bodily and emotional experiences \cite{birdcage}. 

In addition, proactive and instant-reactive features should be designed for the freedom to easily opt-out and opt-in, becoming available to players only when they need it. In Puffer, all features are operated based on voluntary participation since we recognize players' safety needs could change based on the situation. Participants described using Puffer in different scenarios, such as wearing a badge when entering a public game room vs not needing to communicate that preference when hanging out in a private room with friends. Experienced players also shared they would consider blocking all safety suggestions from others because they knew how to activate them and felt confident in locating them in an urgent scenario. 
Voluntary participation engages users who choose to adopt the feature and collectively builds norms in a bottom-up way compared to safety designs that are enforced on individuals through a top-down dictum. 
The enforcement of the Trust Rank system in VRChat, for instance, holds everyone accountable but also poses hierarchical challenges for newcomers and intensifies competition, and researchers suggest that more transparency and engagement between the user and system are needed \cite{vrchatTrust}. \textcolor{edits}{By letting players self-regulate their levels of exposure to potential interactions, we shift the power paradigm to empower each player to take control of their own gaming experiences and, as other researchers identified, continue to support players when other reinforced methods fail \cite{tametoxicplayer}.}

\section{Limitations}
Our recruitment methods relied heavily on online communities and snowball sampling, which led to self-selection bias. \textcolor{edits}{We had a relatively small group of 8 participants for the formative study, with a majority being college students and others recruited through snowballing from VR gaming group chats.} Our participants were all based in the same major city in the U.S. for the in-person study sessions. They were mostly within the same age range (18 to 25) and thus were not representative of the diverse social VR player population. While our recruitment prioritized participants identifying as women or part of the LGBTQ+ community, we did not account for different ethnic and cultural identities. Future research could further our understanding of culture's impact on power dynamics in social VR and how it influences player preferences for safety features. \textcolor{edits}{Additionally, we call for future research to further engage with marginalized players beyond recruitment and create designs tailored to their needs and experiences. }

\textcolor{edits}{We also acknowledge participant response bias in interviews and survey responses, where participants may have been inclined to provide positive feedback to please the researchers \cite{dell_yours_is_better}. We designed ways to mitigate these effects, such as encouraging participants to raise concerns at the beginning of interview sessions. We also asked participants how features could be abused by malicious actors and any scenarios where they would not use the features to seek constructive feedback. Further, we collected data to supplement the self-reported data through observation, such as noting if participants turned on safety features quickly enough before taggers touched them. Despite these efforts, we acknowledge that participant response bias may remain in play.} 

Our simulated scenarios during evaluation cannot replicate a realistic social VR gaming experience, and participants' behaviors in the scenarios were affected by the presence of researchers. Future research with new safety designs in a realistic social VR experience could help us better identify the benefits and challenges of these features in players' daily gaming experiences. \textcolor{edits}{
Since our study focused on three specific design strategies, we lost the opportunity to discover other potential designs. We encourage future work to continue exploring the space of proactive and instant-reactive strategies given their potential to both reduce harm and create positive experiences.}

\section{Future Work}
Participants in our studies suggested ways to improve or expand Puffer. They expressed interest in integrating an automated volume control system for the sound preference badge, allowing noise detection to lower volume when exceeding a standard, particularly addressing screaming, a common form of harassment. Participants also proposed permanent hotkeys for safety features because controller buttons are easier to access than menu navigation. Participants suggested letting players vote on the norms of a game room, involving everyone in a collective process to build norms for the room. 
To extend Puffer, Participants suggested using preference badges to signal specific actions that players like, noting that some types of embodied interactions, such as head pats in VRChat, are key to the gaming experience. 
For the room preference filter, they highlighted the need for labels to warn players with specific triggers (e.g., labeling spiders for people with arachnophobia), as immersive environments could increase discomfort. Future work could explore diverse labeling options to accommodate player needs.

%% file: 8_conclusion.tex
\section{Conclusion}
In this study, we built proactive and instant-reactive strategies to combat embodied harassment in social VR. Proactive designs prevent harassment, and instant-reactive designs provide opportunities for players to intervene in ongoing harassment to minimize harm. Drawing insights from prior work, we approached proactive and instant-reactive designs in three ways: improving user-initiated personal bubbles, clarifying ambiguous social norms, and encouraging bystander intervention. After a formative study exploring the three directions and identifying design goals \textcolor{edits}{, we implemented a social VR safety system called Puffer.} Through evaluation, we found that Puffer fostered more visible social norms, encouraged positive social experiences in social VR, and provided effective protection. We provide insights into designing proactive and instant-reactive features that center marginalized players' strategies and are flexible to adapt to social VR players' evolving and even conflicting safety needs. 


%% file: 9_appendix.tex

\section{Appendix}

\appendix

\section{Landscape Analysis of Personal Bubble}
\label{apx:bubble}
\input{Tables/PB_compare}

To gain a more complete picture of personal bubble across social VR games, we conducted a landscape analysis comparing personal bubble in four popular social VR games---VRChat, Horizon World, RecRoom, and Altspace in October 2022. We identified inconsistencies and potential difficulties in accessing this feature, impacting its effectiveness for protecting players and demonstrating opportunities for improvements. There were four dimensions where personal bubbles were inconsistent across games. First, it was named differently across all games. For example, VRChat names the feature ``Personal Space" while in Horizon Worlds, ``Personal Space" refers to the home space for each player. Second, only RecRoom allows players to customize the size of their bubbles (i.e., small, medium, large). Third, the potential audience of personal bubble varies across games. VRChat and AltSpace provide a binary on/off option. In VRChat, however, even when personal bubble is on, friends can ignore it and enter the player's private space. Horizon Worlds and RecRoom provide the same three choices for players: for everyone, for non-friends, or off. Lastly, the way to access and activate personal bubble varies. It was the easiest to access in AltSpace, as it can be toggled directly from a radial menu. In contrast, other games require players to go through 2-3 steps in the settings page to find their bubble.

\section{Iterations on Safety Suggestions}
\label{apx:iteration}
Participants mentioned safety suggestions being confusing and difficult to understand, and the first prototype has usability issues that limit its ability to be effective in the context of an emergency. Therefore, we made two rounds of iterations on the safety suggestions designs. The first round of changes happened between user study sessions, and the second happened after all sessions completed.

\begin{figure}[h]
    \centering
    \includegraphics[width=0.8\textwidth]{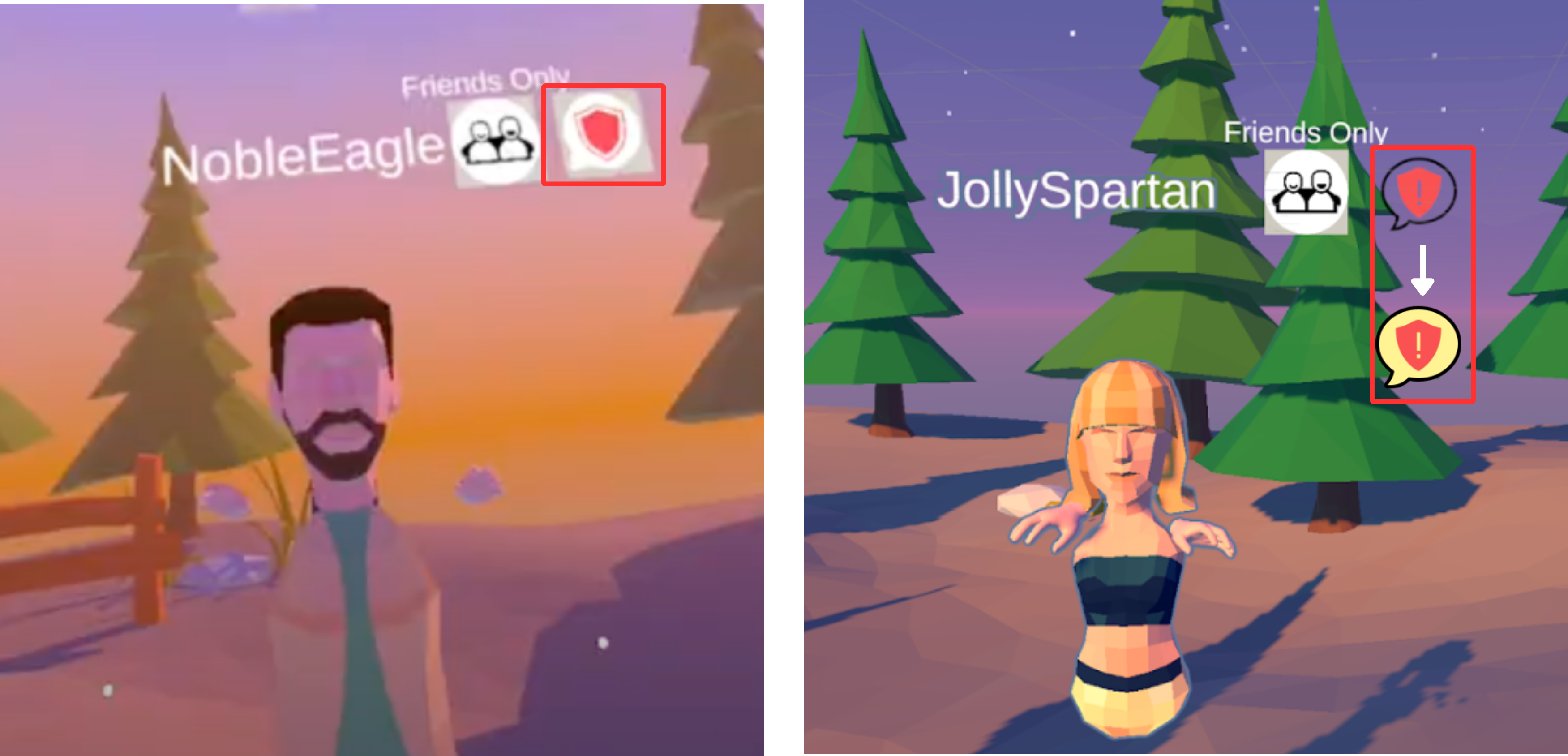}
    \caption{Safety Suggestion Button Design Iteration.}
    \label{fig:ss1}
\end{figure}

First, participants noted that the design of the safety suggestion icon looks too similar to a preference badge, and they did not realize that it was a button they could interact with. We updated the design of the icon and signaled its interactivity by changing the background of the icon to yellow when users' cursor hovers over it, as shown in Figure \ref{fig:ss1}.

\begin{figure}[h]
    \centering
    \includegraphics[width=0.8\textwidth]{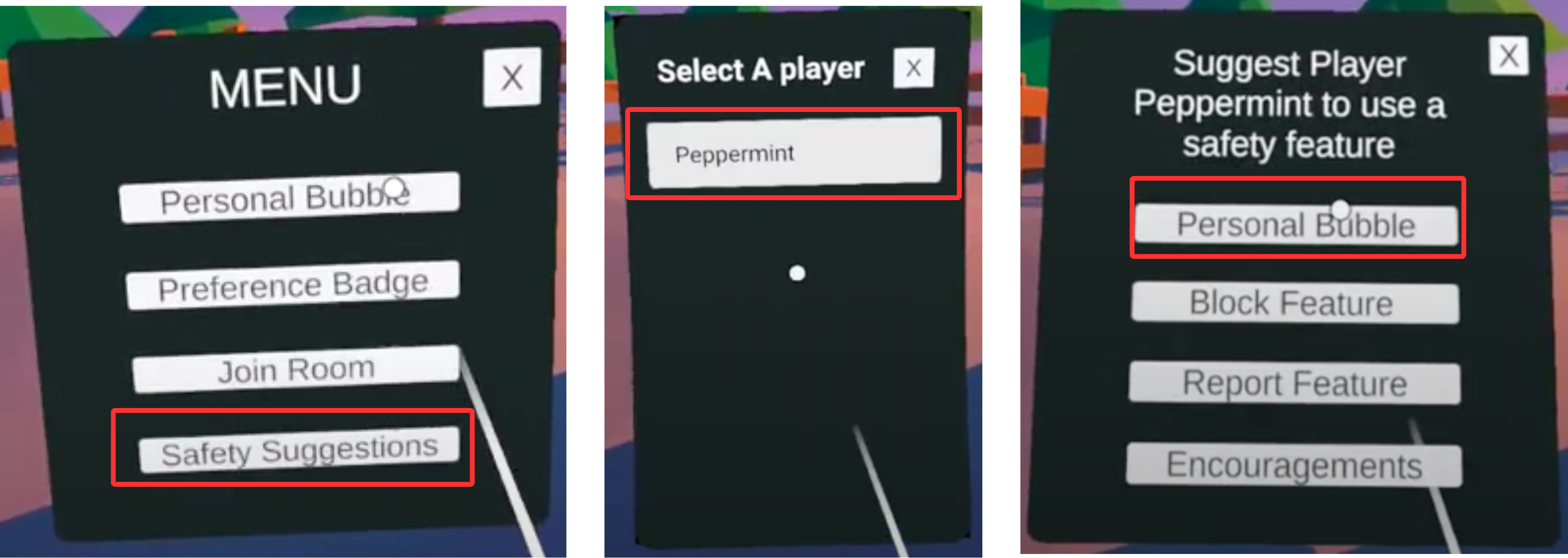}
    \caption{Safety Suggestion Menu Interaction .}
    \label{fig:ss2}
\end{figure}

Second, participants realized that the safety suggestion button is hard to click when the player they want to send a suggestion to is constantly moving, and the button does not support long-distance interaction. We thus added the safety suggestion in the menu system and supported an additional way to send a suggestion, as shown in Figure \ref{fig:ss2}.

From this iteration, we learned that when designing for instant-reactive features, it's important to consider the various conditions players are in an urgent situation, such as their attention or movement, and establish support via multiple pathways.

\section{Formative Study}
\subsection{Design Sketches}
\label{apx:sketches}
During the formative study, we presented design sketches of the bystander intervention design and the preference badge design to participants. These sketches helped illustrate two design directions and elicit participant feedback and thoughts.

\begin{figure}[h]
    \centering
    \includegraphics[width=0.9\textwidth]{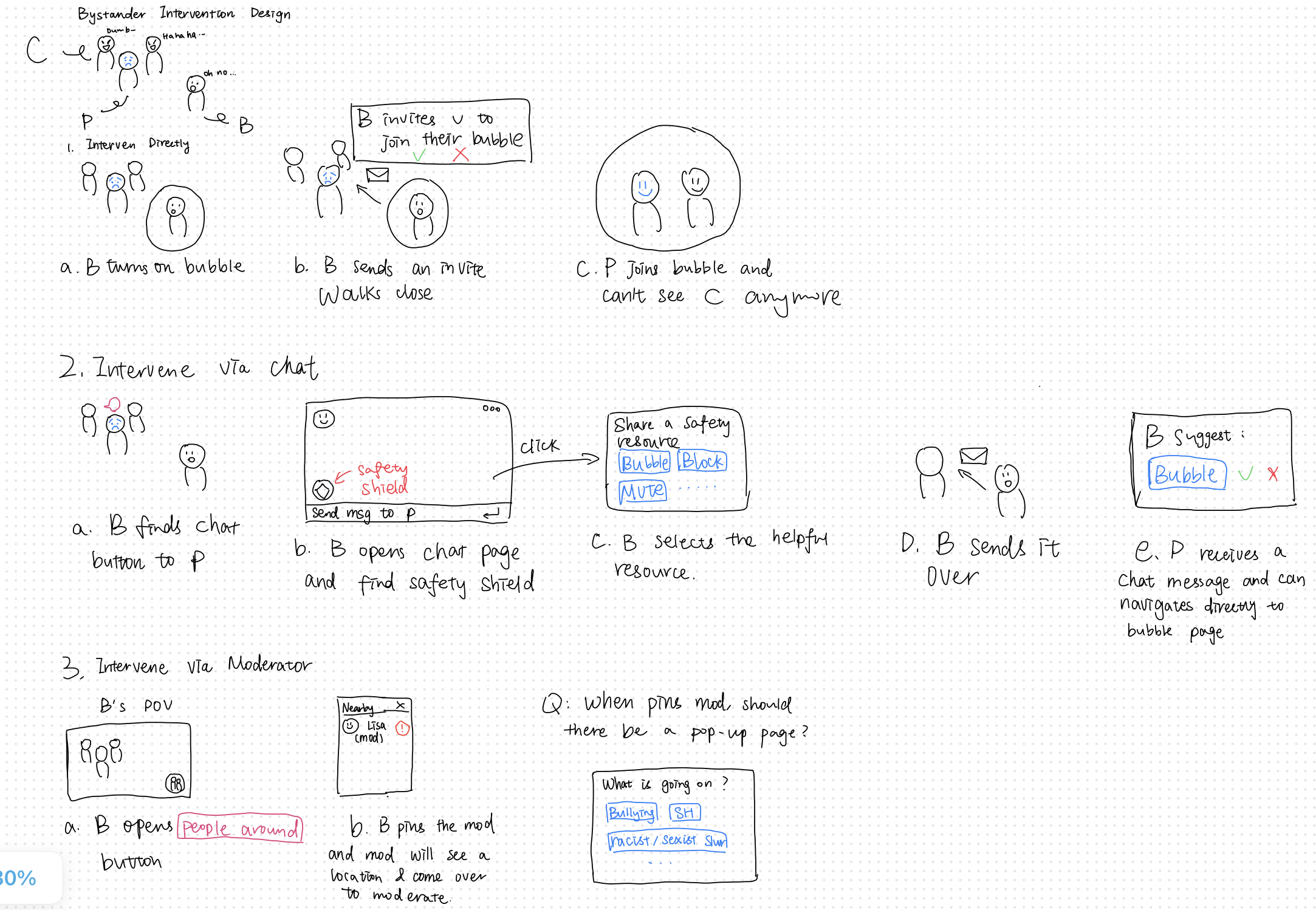}
    \caption{Bystander Intervention Design Sketch. Includes three types of intervention: intervene physically, intervene via chat, and intervene via moderator.}
\end{figure}

\begin{figure}[h]
    \centering
    \includegraphics[width=0.9\textwidth]{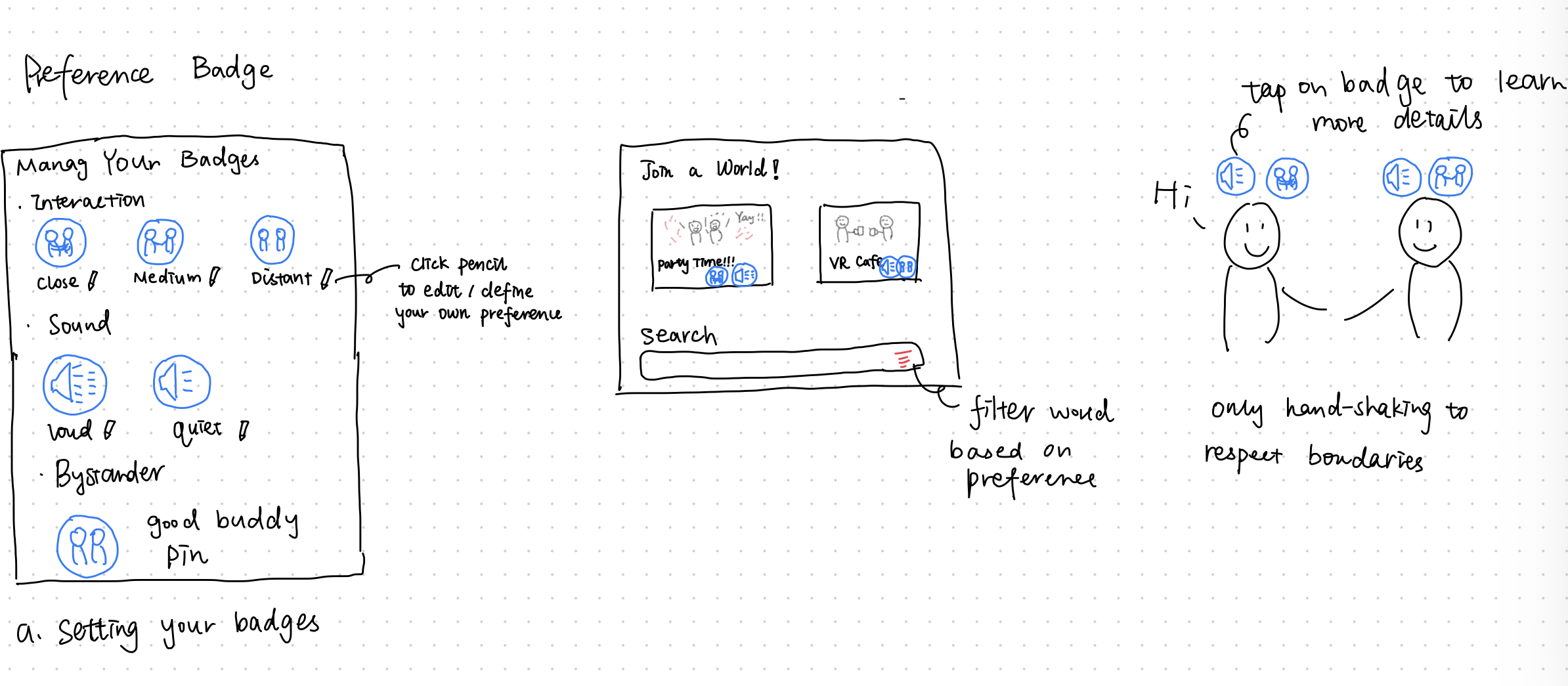}
    \caption{Preference Badge Design Sketch. Include the badge setting interface, filtering game rooms with badges, and badges displaying on top of the avatar's head.}
\end{figure}

\section{Evaluation}
\subsection{Interface Exploration}
\label{apx:interface_exploration}
Participants were given instructions on how to locate each safety feature (Personal Bubble, Preference Badge, Safety Suggestions, Room Preference Filter) in the menus and provided with adequate time to interact with them. Participants were not given any explanations on how to use the feature or the purpose of the feature. They were also instructed to refrain from communicating with each other or asking the researchers any questions. After fully engaging with all four safety features, researchers read four social VR scenario-based tasks to the participants and asked them how they would utilize the safety features in those scenarios Each participant answered two task questions. Researchers also shared that participants are not being tested in any sense, but the system is the one being evaluated and reminded them that they always had the option to reject answering a question, which alleviated stress and encouraged all types of answers and natural reactions from participants. After answering the task questions, participants took off their headsets and debriefed the experience with researchers, such as how intuitive/easy or not to learn each safety feature. 

Here are the four task questions we asked participants to measure learnability of Puffer:
\begin{itemize}
    \item Task 1: You are in a very crowded dancing room and there are so many people close to you. You don’t mind them being “physically” close to you but they are blocking your view. You want to make them invisible to clear your view. What would you do?
    \item Task 2: You are navigating different types of game rooms to see what themes there are. But you are not in the mood to talk to anyone (you want to be alone) and you want to indicate this mood to other players so they don’t attempt to talk to you. What would you do? 
    \item Task 3: You want to explore some new game rooms in VR. You want to find a room that is not crowded or noisy. What would you do?
    \item Task 4: You saw one group of people making fun of another player. You want to help the player and suggest the player to both turn on the personal bubble and block them. What would you do?
\end{itemize}

This scenario mimics the situation where players try to figure out how to use and understand the effect of safety features solely by themselves, where they lack assistance from outsiders such as friends or moderators and can only depend on signifiers built in the system and information on the user interface. This is a common situation for all new players entering the social VR game who do not have any friends yet. Although tutorials are necessary for players to learn about safety features suggested by the formative study, it is, however, frequent that players tend to ignore or skip any tutorials. Thus, this scenario is designed to occur before participants watch the tutorial. This scenario evaluates how easily participants can learn to use the system with minimum assistance. 

\subsection{Simulated Scenarios}
\subsubsection{Roleplay Scenario}
\label{apx:roleplay}
Before entering the second scenario, participants watched a tutorial video explaining all safety features, which is a similar video mentioned in Chapter 4, and had opportunities to propose any questions about the system. This ensures that participants continued scenarios 2 and 3 with adequate knowledge about the system. In the roleplay scenario, participants were randomly assigned a persona written on a card that represents a social VR player. The persona card includes information about the motivation and priorities in a social VR game and their preferences. Persona 1 is a nervous new player, Peppermint, and Persona 2 is a friendly, experienced player, Thunderhawk. These two personas present opposite priorities and preferences in a social VR game where Peppermint is cautious and avoids interactions, and Thunderhawk wants to interact with players to make friends. Participants were asked to act according to their persona throughout the whole scenario. They were also asked not to reveal their persona to the other participants in the section. After acting through all scenarios, participants took off their headsets and debriefed the experience of interacting with other players in the game while adopting safety features.

This scenario is designed to simulate a real interaction between players in a social VR game in an experimental setting. By introducing safety features in the tasks, we observe and comprehend these features' influence on players' perception of other players, and thus their reactions and behaviors. 

Here are the two personas that participants were randomly assigned to for roleplay:
\begin{itemize}
    \item \textbf{Nervous New Player - Player name: Peppermint} This is your first time playing Social VR. You are navigating a public game lobby room and feeling very nervous. 
    \begin{itemize}
        \item You do NOT want to have any “physical” interaction with strangers.
        \item You do NOT want other players to get close to you.
        \item You do NOT know anyone in the game and you are very cautious of strangers. 
        \item You prefer to be ALONE
        \item You are OKAY with talking to strangers if they seem friendly
    \end{itemize}
    \item \textbf{Friendly Experienced Player - Player name: Thunderhawk} You are an experienced player in this game. You are familiar with all the safety features. You are friendly and willing to help others. You want to make new friends through this game. If possible, you want to find game rooms that have a lot of players so you can make more friends. You want to make friends with player Peppermint because you like their avatar and they seem cool
\end{itemize}

During scenario 2, participants roleplayed the following tasks with prompts from researchers:
\begin{enumerate}
    \item \textbf{Social Interaction A: } Without any safety features, Thunderhawk tries to make friends with Peppermint.
    
    In this task, both participants turn off their personal bubble and wear no badge, having no indicators of their preferences. When Thunderhawk approaches Peppermint, an alert message pops up on the screen for both participants to remind them someone is close by. Participants were observed on their reactions to these alert messages (for example, if Peppermint took action to protect themselves when being approached by a stranger) and how they navigated a social interaction without understanding other players' preferences and intentions.
    
    \item \textbf{Social Interaction B: } Both participants wear preference badges to indicate their priorities. Thunderhawk tries to make friends with Peppermint.
    
    Task 2 replays task 1 with the distinction that both participants use preference badges. Participants were observed interpreting each other's preferences with the badge and their actions. They also shared their thought processes during these two tasks when debriefing.
    
    \item \textbf{Room Finding: } Both participants try to find a room that matches their interaction and sound preference.
    
    This task enables participants to utilize the room preference filter feature with explicit user motivations to mimic a real usage of this feature if implemented in a social VR game. 
    
    \item \textbf{Bystander Intervention:} Two researchers put on Oculus Quest2 headsets and join the prototype in the same game room as the participants. One researcher's avatar closely chases another researcher's avatar, and both remain silent in the process. Participants observe the researchers' actions and react based on their personas. 
    
    In the last scenario, participants were placed as bystanders to observe researchers. Intentionally, researchers simulate a potential embodied harassment situation without clear cues such as the language of conflicts or hate. This avoids directly placing participants in harmful or triggering positions and more truthfully represents the ambiguity of harassment. As suggested during the formative study (add a reference here), participants found it difficult to distinguish between harassment and friends messing around if no obvious cues were presented. During this scenario, we observed how participants judged situations of potential harassment and how they reacted or intervened. Additionally, we included the preference badge as a variable among all sessions. For half of the sessions, the researcher who was chased wore an individual social energy badge and 'No physical interaction' badge and did not wear anything in the other half. Our hypothesis behind this manipulation is that the presence of a preference badge helps players interpret the situation better and intervene in the harassment. 
\end{enumerate}

\subsubsection{Tagging Game}
\label{apx:tag}
Participants played a tagging game with the researchers in a Halloween-themed game room within the prototype, where the researchers acted as the taggers and participants avoided being tagged. A participant was considered "tagged" if the taggers' avatar hands touched their avatar bodies. Participants could turn on their personal bubble to guard their personal space and avoid being tagged. They were also assigned a counting task to perform during the game to serve as a distraction. However, since turning on the personal bubble would make them safe and thus end the game, we asked the participants to focus on the task most of the time and refrain from activating the personal bubble unless the tagger was close enough that they perceived danger. The game was played three times, each offering a different way for the player to access the personal bubble, with the order of these variations randomized to mitigate any ordering effects. 
\begin{itemize}
    \item Setup 1: Activate the personal bubble by navigating through the menus and clicking the check box. 
    \item Setup 2: Receiving an alert message about players close by and pressing the shortcut button on the controller.
    \item Setup 3: One participant acted as the helper and one as the player. The player activated the personal bubble by receiving a safety suggestion from the helper. 
\end{itemize}
We recorded the results of whether players successfully avoided being tagged. Participants debriefed their experiences of accessing the personal bubble in various methods. 

This scenario simulates emergency situations where players perceive potential discomfort or danger or are already experiencing harassment and need to activate the safety features immediately to stop future harm. We designed a tag game to replicate the sense of urgency when embodied harassment happens in a fun and lighthearted way, avoiding direct exposure to a real harassment situation that could be unpleasant or triggering for participants. By conducting the tag games, we compare multiple ways to access the safety features and understand their effectiveness in the context of an emergency.

\subsection{Interview and Survey Questions}
\label{apx:questions}
There were five interview questions asked for each safety feature in Puffer:
\begin{enumerate}
    \item Would you consider using [feature] in Social VR games? If yes, in what contexts? If no, why?
    \item How effective do you think [feature] is at protecting players’ safety? What about under emergency situations?
    \item What impact do you think [feature] has on the norms and dynamics in the game? 
    \item How concerned are you about [feature] being abused by malicious individuals?
    \item What suggestions do you have for improving this [feature]?
\end{enumerate}

Exit survey questions was split into four sections, one for each safety feature. Each section contains the same set of 10 questions from the SUS where the participants rated from Strong Disagree to Strong Agree:
\begin{enumerate}
    \item I think that I would like to use this feature frequently.
    \item I found the feature unnecessarily complex.
    \item I thought the feature was easy to use.
    \item I think that I would need the support of a technical person to be able to use this feature.
    \item I found the various functions in this feature were well integrated.
    \item I thought there was too much inconsistency in this feature.
    \item I would imagine that most people would learn to use this feature very quickly.
    \item I found the feature very cumbersome to use.
    \item I felt very confident using the feature.
    \item I needed to learn a lot of things before I could get going with this feature.
\end{enumerate}

Additionally, each feature had some unique questions about its effectiveness:
\begin{itemize}
    \item \textbf{Personal Bubble}
    \begin{enumerate}
        \item How helpful Personal Bubble is in protecting players from embodied harassment?	\item How helpful Personal Bubble is under emergency situations?
        \item How helpful it is to have Hard Boundary in game?
        \item How helpful it is to have Soft Boundary in game?
        \item How helpful the alert message (reminding players to turn on Personal Bubble) is?
    \end{enumerate}
    \item \textbf{Preference Badge}
    \begin{enumerate}
        \item How helpful badges are in supporting you to understand other players’ preferences?
        \item How helpful badges are in communicating your preferences with other players?
        \item How helpful badges are in building better social norms in game?
        \item I am concerned that this feature will be abused by malicious individuals. 
    \end{enumerate}
    \item \textbf{Safety Suggestion}
    \begin{enumerate}
        \item How helpful Safety Suggestion is in locating safety features?
        \item How helpful Safety Suggestion is in encouraging players to mutually help each other?
        \item How helpful Safety Suggestion is in building better social norms in game?
        \item I am concerned that this feature will be abused by malicious individuals. 
    \end{enumerate}
    \item \textbf{Room Preference Filter}
    \begin{enumerate}
        \item How helpful Room Preference Filter is in finding a room that matches your preferences?	
        \item How helpful Room Preference Filter is in building better social norms in game?
        \item I am concerned that this feature will be abused by malicious individuals. 
    \end{enumerate}
\end{itemize}

\subsection{Survey Results}
\label{apx:ratings}
\textcolor{edits}{
In the exit survey, participants were neutral about the features being abused. Between 1 to 5, 1 being not concerned at all and 5 being extremely concerned, preference badges scored an average of 3.13 across all 16 participants, safety suggestions scored 3.31, and the room preference filter scored 2.25.\\
Usability was measured using the SUS in the exit survey. The personal bubble scored 75.71, preference badges scored 83.13, safety suggestions scored 67.03, and room preference filters scored 85.31.}

\section{Codebook}
\subsection{\textcolor{edits}{Formative Study Codebook}}
\label{formative_codebook}
\input{Tables/formative_study_codes}

\subsection{\textcolor{edits}{Evaluation Codebook}}
\label{final_codebook}
\input{Tables/evaluation_codes}

%% file: Tables/PB_compare.tex
\begin{table}[H]
\small
\centering
\begin{tabular}{lllll}
\hline
\multicolumn{1}{|l|}{} &
  \multicolumn{1}{l|}{\textbf{Name}} &
  \multicolumn{1}{l|}{\textbf{Size}} &
  \multicolumn{1}{l|}{\textbf{Audience}} &
  \multicolumn{1}{l|}{\textbf{Access}} \\ \hline
\multicolumn{1}{|l|}{\textbf{VRChat}} &
  \multicolumn{1}{l|}{Personal Space} &
  \multicolumn{1}{l|}{Not Specified} &
  \multicolumn{1}{l|}{Only Non-Friends} &
  \multicolumn{1}{l|}{QuickMenu -\textgreater Settings} \\ \hline
\multicolumn{1}{|l|}{\textbf{\begin{tabular}[c]{@{}l@{}}Horizon\\ Worlds\end{tabular}}} &
  \multicolumn{1}{l|}{Personal Boundary} &
  \multicolumn{1}{l|}{4 Feet} &
  \multicolumn{1}{l|}{\begin{tabular}[c]{@{}l@{}}Everyone or Non-\\ Friends\end{tabular}} &
  \multicolumn{1}{l|}{Menu -\textgreater Settings -\textgreater Safety} \\ \hline
\multicolumn{1}{|l|}{\textbf{Altspace}} &
  \multicolumn{1}{l|}{Personal Space Bubble} &
  \multicolumn{1}{l|}{Not Specified} &
  \multicolumn{1}{l|}{Only Everyone} &
  \multicolumn{1}{l|}{Radial Menu} \\ \hline
\multicolumn{1}{|l|}{\textbf{RecRoom}} &
  \multicolumn{1}{l|}{\begin{tabular}[c]{@{}l@{}}Personal Space Bubble \\ and Ignore Bubble\end{tabular}} &
  \multicolumn{1}{l|}{S, M, L} &
  \multicolumn{1}{l|}{\begin{tabular}[c]{@{}l@{}}Everyone or Non-\\ Friends\end{tabular}} &
  \multicolumn{1}{l|}{\begin{tabular}[c]{@{}l@{}}Wrist Menu -\textgreater Settings -\textgreater \\ Experience\end{tabular}} \\ \hline
\end{tabular}
\caption{Comparison of Personal Bubble Across Social VR Games. Note: Audience column refers to whom are under the effects of the bubble.}
\label{tab:bubble}
\end{table}

%% file: Tables/formative_study_codes.tex
\begin{table}[h]
\centering
\begin{tabular}{>{\color{edits}\arraybackslash}l >{\color{edits}\arraybackslash}l}
\toprule
Theme & Code \\
\midrule
Lack of knowledge about personal bubble	& Not aware of existing safety features, e.g. personal bubble \\
& Need more visualization to understand the size of bubble\\
\midrule
Hard to access personal bubble & Difficult to find the personal bubble from the menu \\
& Settings menu titles are confusing \\
& Not useful for emergent situations \\
\midrule
Preferences \& challenges for social interactions in game & Want to signal social intent to others \\
& Do not want to interact with strangers in social VR \\
& Do not share interaction preference with others \\
& Do not know how to communicate interaction preference \\
& Filter is helpful for finding the ideal game world \\
\midrule
Preferences \& challenges for managing conflicts & Lack of trust for moderators \\
& Leverage friends for support \\
& Rely on oneself to manage conflicts/harassment \\
& Give players agency to protect themselves \\
& Avoid getting personally involved or escalation \\
& Bystander's concern about misreading the situation \\
\midrule
Concern about abusing safety features & Lack of trust in strangers in social VR/zero trust for strangers \\
& Spam others with safety chat messages \\
& Badge turns players into targets \\
\midrule
Suggestions \& Improvements	& Badge requires many people's participation \\
& Prefer a bigger default size of personal bubble \\
& Users prefer ways to customize the effect of personal bubble \\
& Badge could be distracting \\
\bottomrule
\end{tabular}
\caption{Codebook for the inductive thematic analysis in the formative interview study.}
\label{tab:formative-codebook}
\end{table}

%% file: Tables/evaluation_codes.tex
\begin{table}[h]
\centering
\begin{tabular}{>{\color{edits}\arraybackslash}l >{\color{edits}\arraybackslash}l}
\toprule
Theme & Code \\
\midrule
Protect players' safety & \\
- Fast and easy access to safety features & hotkeys are quick shortcuts to access safety features \\
& Help new players to get safety support more easily and quickly \\
- Protect users from engaging in unwanted situations & Avoid unwanted physical interactions \\
& Avoid escalating players' conflicts \\
& Increase awareness about players around me \\
- Increased context promote safe interactions & Improved transparency in context understanding to support others \\
& Prevent unintentional conflicts or disrespect \\
\midrule
Clarify social norms & Signaling expectations of interacting with another individual \\
& Players respect others' interaction preferences \\
& Clarifying norms/expectations for a virtual world \\
\midrule
Improved social experience & Mutual help lead to positive social experience \\
& Help other players without endanger oneself \\
& Encourage players to look out for each other \\
& Find players who share a similar vibe \\
& Find virtual worlds that match the energy \\
& Soft boundary could decrease crowdedness/visual noise \\
& Help new players to make friends more easily \\
\midrule
Abuse of system	& Visible personal preferences become targets \\
& Intersectional impact of social badge \\
& Room labels become targets \\
& Spam safety suggestions \\
\midrule
Misinterpretation of safety suggestions	& Suggestions received as passive-aggressive \\
& Create doubt among users \\
& Suggestions are lack of specific contexts \\
& Misread the intention of safety suggestions \\
\midrule
Reduced/Impacted User Experience & Limit the physical interaction/movement \\
& Increase stress or overwhelmingness during emergencies \\
& Seeing people running into the hard boundary \\
& Detection messages distract players when pop up too often \\
\midrule
Suggestions \& Improvements & Control/filter auditory noise \\
& Combine personal badges with bubbles \\
& Change the wording for badges \\
& Allow room filters for popular rooms \\
& Extend the room filters for phobias \\
\midrule
Other \& Miscellaneous & Generalization for other moba games \\
& Features mimic real-world examples \\
\bottomrule
\end{tabular}
\caption{Codebook for the inductive thematic analysis in the evaluation.}
\label{tab:evaluation-codebook}
\end{table}